\RequirePackage{lineno}
\documentclass[reprint,prd,tightenlines]{revtex4}

\usepackage{graphicx} 
\usepackage{dcolumn}  
\usepackage{colordvi}
\usepackage{color}
\usepackage{epstopdf}
\usepackage{amssymb}
\usepackage{url}
\graphicspath{{ps}}
\usepackage{hyperref}
\usepackage{tabularx}
\usepackage{lineno}
\usepackage{multirow}
\usepackage{longtable}
\usepackage{appendix}
\usepackage{adjustbox}
\usepackage{amsmath}
\usepackage{enumitem}
\usepackage{tabularx}

\makeatletter

\newcommand{\Rmnum}[1]{\expandafter\@slowromancap\romannumeral #1@}
\makeatother

\newcommand{\be}{\begin{equation}}
\newcommand{\ee}{\end{equation}}
\newcommand{\ba}{\begin{array}}
\newcommand{\ea}{\end{array}}
\newcommand{\bqa}{\begin{eqnarray}}
\newcommand{\eqa}{\end{eqnarray}}
\newcommand\tbbint{{-\mkern -16mu\int}}\newcommand\dbbint{{-\mkern -19mu\int}}\newcommand\bbint{{\mathchoice{\dbbint}{\tbbint}{\tbbint}{\tbbint}}}

\begin{document}

%
%

\title { \quad\\[0.5cm] Practical parametrization of two-pole structure}

\author{J.X. Cui}
\affiliation{School of Physics, Southeast University, Nanjing 211189}

\author{Zhiguang Xiao}
\email{xiaozg@scu.edu.cn}
\affiliation{Institute for Particle and Nuclear Physics, College of Physics, Sichuan University, Chengdu  610065, P.~R.~China}

\author{Zhi-Yong Zhou}
\email{zhouzhy@seu.edu.cn}
\affiliation{School of Physics, Southeast University, Nanjing 211189}

\begin{abstract}


We suggest that the extended Lee-Friedrichs model could be directly used as a practical parametrization method for the experimental analysis of resonance structures. This parametrization  incorporates the constraints of relativistic phase space and the threshold behavior, and respects both unitarity and analyticity constraints of the scattering amplitude. As such, the poles on unphysical Riemann sheets could be easily extracted.  This parametrization method  offers a comparable fit quality to the improved Breit-Wigner parametrization with an energy-dependent width function when the coupling strength is moderate. It is found that the parametrization could be used to correctly extract the poles near the physical region correctly. In particular, it can naturally incorporate the two-pole structure in which one pole is shifted from the discrete state and the other is dynamically generated.
Moreover, the coupled-channel formulation of the extended Lee-Friedrichs parameterization is straightforward and its relationship with the Flatt\'{e} parametrization form is discussed. Using $\rho(770)$, $\Delta(1232)$, $K^*_0$ and $f_0$ states as illustrative examples, we demonstrate the effectiveness of this parametrization in capturing fit qualities and identifying relevant poles. It is illustrated that the $K_0^*(700)$ and $K_0^*(1430)$ could be perfectly parameterized in the Lee-Friedrichs form as a two-pole structure. A tentative investigation of $f_0$s  in coupled-channel parametrization form are discussed, and a possible lineshape contributed by $\chi_{c1}(3872)$ and $\chi_{c1}(4012)$ is presented.
The proposed parametrization scheme holds promise for future studies involving exotic hadron states near thresholds, offering a valuable tool for analyzing resonance structures in upcoming experimental investigations.

\end{abstract}


\maketitle

{\renewcommand{\thefootnote}{\fnsymbol{footnote}}}
\setcounter{footnote}{0}

%
\newpage

\section{Introduction \label{introductionSection}}

In scattering theory, a resonance state is typically associated with a pole located on the complex $s$ plane of unphysical Riemann sheets of the scattering amplitude, with $s$ being the usual Mandelstam variable. In the context of an isolated narrow resonance state located away from thresholds, experimental observations typically exhibit characteristic features in the cross section data, such as peaks, kinks, or dips~\cite{Taylor:1972pty}. Traditionally, these resonance structures are commonly described and parameterized using the Breit-Wigner (BW) distribution. With this distribution, the partial wave scattering amplitude near the resonance is described as
\be
A\propto \frac{M_{BW}\Gamma_{BW}}{M_{BW}^2-s-iM_{BW}\Gamma_{BW}}\simeq \frac{\Gamma_{BW}/2}{M_{BW}-E-i\Gamma_{BW}/2}
\ee
which corresponds to a resonance with the mass $M_{BW}$ and the width $\Gamma_{BW}$ where $E\equiv \sqrt{s}$.  When the higher resolution data are obtained and the resonance width is not small, the production amplitude of BW parametrization observed in channel $a$ is improved to fit data by introducing an energy-dependent width as
\bqa\label{improvedBW}
A_a\propto \frac{N_a(s)}{M_{BW}^2-s-iM_{BW}\Gamma(s)}.
\eqa
The polynomial functions $N_a(s)$ and $\Gamma(s)$ are defined as
\bqa
N_a=\alpha g_a n_a(s), \ \ \Gamma(s)=\frac{1}{M_{BW}}\sum_b g_b^2\rho_b(s)n^2_b(s)
\eqa
where the index $a$ and $b$ denotes the scattering channels coupled with the resonance, $g_a$ and $g_b$ the coupling constants, and $\rho_b(s)$ the phase space factor of channel $b$. The expressions for $n_a(s)$ and $n_b(s)$ factors are required to satisfy the threshold behavior constraints and adhere to the Blatt-Weisskopf barrier factor phenomenologically. This improved BW parametrization method with energy-dependent width is widely recognized and further details can be found in the review by the Particle Data Group (PDG) \cite{ParticleDataGroup:2022pth}.

While the BW parametrization remains prevalent in experimental data analysis, it suffers from several well-documented uncertainties. One significant issue is the ambiguity of multiple solutions: fitting procedures may yield various solutions of comparable quality but with different physical implications when multiple resonances sharing the same quantum numbers are present in the same energy range \cite{Zhu:2011ha, Bukin:2007kx, Han:2018wbo, Bai:2019jrb}.
Another source of uncertainties stems from violations of unitarity. In scenarios involving neighboring interfering resonances, the amplitude described by the sum of multiple BW resonances may fail to satisfy the constraint of partial-wave unitarity.
To address the limitations of the BW parameterization, alternative methods such as $K$-matrix, Flatt\'{e}, and other formulations have been developed. More recently, additional new parametrization techniques have been introduced~\cite{Giacosa:2021mbz,Heuser:2024biq}. { These approaches advocate for the assignment of one pole to one form, akin to the practice seen in BW parameterization.}

In the extensive exploration of hadron physics, both from theoretical and experimental perspectives, there is a growing number of instances where two hadronic states may be related. Frequently, one of the states is easy to identify but the other is hard to be determined, as its lineshape deviates from the typical description provided by BW parametrization. Consequently, such states often arouse controversies until more robust model-independent theoretical analyses, respecting basic principles like unitarity and analyticity, confirm their existence. A prominent example of this phenomenon is observed in the cases of the $f_0(500)$ and $K_0^*(700)$ states. Recent investigations indicate that some of these states may not conform to conventional quark model descriptions but are potentially accompanying with other higher conventional states due to the strong couping between the conventional states and the scattering channels~\cite{Zhou:2020moj,Zhou:2020vnz,Eichten:1979ms, vanBeveren:1983td, Tornqvist:1995kr, Li:2009pw, Limphirat:2013jga, Achasov:2012ss, Zhang:2009gy, Segovia:2011zza, Wolkanowski:2015lsa, Wolkanowski:2015jtc, Yao:2020bxx}. This two states appear together as two pairs of poles of the $S$-matrix and can be effectively described by the extended version of the Lee-Friedrichs~(LF) model \cite{Zhou:2020moj,Zhou:2020vnz,Lee:1954iq,Friedrichs:1948}. Consequently, this configuration is termed the two-pole structure. From our perspective, such two-pole structures are quite prevalent in the hadron spectrum due to the strong interaction of the discrete states and the channels they couple to. This is a result of the fact that the dressed propagator often exhibits multiple poles~\cite{Zhou:2010ra,Zhou:2020moj,vanBeveren:1986ea,Tornqvist:1995kr,Boglione:1997aw,Giacosa:2019ldb}. In certain cases, only one pole lies in close proximity to the physical region such that it can be readily identified from the experimental data, making it  amenable to parameterization using the BW form and  easy to be extracted. In these scenarios, the contribution of the other poles  exhibit a mild smooth structure that can be assimilated into the background contribution, which is always a conventional assumption in experimental analyses.
However, complications arise when the two poles are close to each other and are hard to extract simultaneously without additional assumptions about the general properties of the scattering amplitude. It is crucial to note that the distinct lineshape peaks observed in such two-pole structures. Parameterizing each pole separately with a BW form leads to violations of unitarity and introduces unnecessary parameters, potentially diminishing the significance of the states. Additionally, when one of the poles represents a dynamically generated bound state just below the threshold and its lineshape is distorted by the threshold, it would be very difficult to use the BW parametrization to accurately determine the mass and width parameters.

In this study, we introduce a parameterization form derived from the LF model as a single parameterization form to characterize the two-pole structure in analyzing the experimental data. This parametrization form inherently respects partial-wave unitarity and analyticity, facilitating the straightforward extraction of the poles.
In the basic framework of the LF parametrization involving only one bare state coupled to a single decay channel, it is common to find that two pairs of complex poles on the complex energy plane of unphysical Riemann sheets contribute to the observables. These pole pairs are typically interpreted as two distinct states within the experimental analysis, while in the proposed parameterization they are correlated with each other with fewer free parameters.

The paper is organized as followings: the parameterization method is introduced in Section~\ref{paraSection}. The analytical continuation and the pole extraction are discussed in details in Section~\ref{analyticcontinuation}.  Section~\ref{numerical} presents the detailed analyses on several sets of example experimental data using the LF parametrization, and we provide some discussions and summarize the work in Section~\ref{summarySection}.

\section{Parameterization Method \label{paraSection}}

In this section, we briefly introduce the LF model and discuss the parameterization method in detail.

	\subsection{General Scheme}

The original LF model is a solvable model used to study the generating mechanism of resonance when one bare state interacts with one scattering channel~(or referred to as a continuum state)~\cite{Lee:1954iq,Friedrichs:1948}.	From the ``in" and ``out"  scattering state solutions, one can derive the scattering amplitude, which satisfies the unitarity and analyticity prerequisites that are essential for elastic scatterings. It is found that generally the bare state is shifted to the complex energy plane and some extra poles could be dynamically generated as the coupling strength increases~\cite{Xiao:2016dsx,Xiao:2016wbs,Likhoded:1997}.

When extending the model to a more general scenario for practical applications, we can consider a system in the center of mass system that comprises $N_d$ types of discrete states and $N_c$ types of scattering channels they couple to, where $N_d$ and $N_c$ represent the respective counts of bare discrete states and scattering channels. In the absence of interactions, the mass of the $a$-th discrete state $|a\rangle$ is denoted by $M_a$, while the threshold energy of the $j$-th scattering channel $|E;j\rangle$ is represented by $E_j$. The interaction between the $a$-th discrete state and the $j$-th
scattering channel   can generally be represented by a coupling function $f_{ja}(E)$ . The full Hamiltonian can be expressed as
\begin{align}
H=&H_0+H_I,
\end{align}
where the free Hamiltonian $H_0$  could be written down explicitly as
\begin{align}
H_0=&\sum_{a=1}^{N_d} M_a|a\rangle\langle
a|+\sum_{j=1}^{N_c} \int_{E_j}^\infty \mathrm d E
\,E|E;j\rangle\langle E;j|,
\end{align}
and the interaction part $H_I$ reads
\begin{align}
H_I=&\sum_{a=1}^{N_d}\sum_{j=1}^{N_c} \left[ |a\rangle\Big(\int_{E_j}^\infty\mathrm d E
f^*_{ja}(E)\langle E;j|\Big)+ \Big(\int_{E_j}^\infty\mathrm d E
f_{ja}(E)|E;j\rangle \Big)
\langle a|
\right].
\end{align}
The scattering matrix element for the amplitude of the $j$-th channel to the $i$-th channel can be written down as~\cite{Xiao:2016mon,zhou:2020,Xiao:2023lpv}
\begin{align}\label{eq:SmatrixNR}
{\cal{S}}_{ij}(E) = \delta_{ij}-2\pi i\sum_{a,b}^{N_d}f_{i a}(E)[\eta^{-1}]_{ab}f^*_{j b}(E),
\end{align}
for a non-relativistic scenario,
where $i$ and $j$ are the indices of scattering channels, $a$ and $b$ indices of the discrete states. $[\eta]$ being the inverse resolvent function matrix of dimension $N_d\times N_d$ reads
\begin{equation}
[\eta]_{ab}\equiv(E-M_a)\delta_{ab}-\sum_{i}^{N_c}\int_{{E_{i}}}^{\infty} d E^{\prime}\frac{f_{i a}(E^{\prime})f^*_{i b}(E^{\prime})}{E-E^{\prime}}.
\end{equation}

This model can be generalized to describe the relativistic systems by including some relativistic considerations. The partial wave scattering amplitude could be derived by diagonalizing the four-momentum operator $P_\mu$, as depicted in ref.~\cite{zhou:2020}. The result is similar to the unitary quark model where the scattering amplitude is directly written down by assuming the coupled-channel unitarity and dispersion relation~\cite{Tornqvist:1995kr}. The  resultant partial wave $S$-matrix is
\begin{align}\label{eq:SmatrixR}
{\cal{ S}}_{ij}(s) = \delta_{ij}-2\pi i\sum_{a,b}^{N_d}f_{ia}(s)[\eta^{-1}]_{ab}f^*_{jb}(s).
\end{align}
which is akin to the non-relativistic version.
Here the Mandelstam variable $s$ represents the squared energy of center-of-mass (c.m.) system.
The inverse of the resolvent matrix $\eta$ reads
\begin{equation}\label{etafunc}
[\eta]_{ab}\equiv(s-M_a^2)\delta_{ab}-\sum_{i}^{N_c}\int_{{s_{th,i}}}^{\infty} d s^{\prime}\frac{f_{ia}(s^{\prime})f^*_{ib}(s^{\prime})}{s-s^{\prime}},
\end{equation}
where  ${s_{th,i}}$ is the threshold energy squared for $i$-th scattering channel.
To make the representation simpler, we can define the spectral density function $\sigma_{ab,i}(s)\equiv f_{ia}(s)f^*_{ib}(s)$, and the self-energy function is expressed as
\begin{equation}\label{eq:selfenergy}
\Pi_{ab,i}(s)=\int_{{s_{th,i}}}^{\infty} d s^{\prime}\frac{\sigma_{ab,i}(s^{\prime})}{s-s^{\prime}}.
\end{equation}
Then $\eta$ function can be represented as
\begin{equation}\label{eq:eta}
[\eta]_{ab}=(s-M_a^2)\delta_{ab}-\Pi_{ab}(s),
\end{equation}
where $\Pi_{ab}(s)\equiv\sum_{i=1}^{N_c}\Pi_{ab,i}(s)$.

The $\eta$ functions  in both the non-relativistic and  relativistic scenarios exhibit branch cuts extending from the thresholds to the infinity, and the partial wave $S$-matrix satisfy the partial wave unitarity and analyticity, which can be readily verified.
By analytically continuing the $\eta(E)$ or $\eta(s)$  functions along the unitarity cut to the complex $E$ or $s$ plane respectively, zero points of $\mathrm{det}\eta$ emerge on different Riemann sheets, corresponding to the poles of the $S$-matrix which represent the resonance states. It is important to note that an infinitesimal imaginary part of energy, $E+i0$ or  $s+i0$ is implicitly included when the eqs.(\ref{eq:SmatrixNR}) and (\ref{eq:SmatrixR}) are used to represent the $S$-matrix element within the physical region.

\subsection{The  parameterization method}
We are going to illustrate the parameterization method starting from the simple case to the complex ones. Each case is denoted by the numbers of the discrete states $N_d$ and scattering channels $N_c$ as $(N_d,N_c)$.  In the simplest case $(1,1)$, the method is elaborated in both the non-relativistic and relativistic scenarios. For other cases, only the relativistic version is discussed, and the extension to the non-relativistic version is straightforward.	

\subsubsection{The $(1,1)$ case}
		
The simplest case involves one discrete state coupling with a two-body scattering channel, where the indices of the discrete state and scattering channel can be omitted.

\emph{Relativistic scenario}

In this case, the $S$-matrix is very simple and expressed as
\begin{align}
\mathcal{S} (s)   &=1-\frac{2\pi i \sigma(s)}{s-m^2-\Pi(s)},
\end{align}
where $\Pi(s) =\int_{s_{th}}^{\infty}d s^{\prime}\frac{\sigma(s^{\prime})}{s-s^{\prime}}$ and $s_{th}=(m_1+m_2)^2$  denoting the masses of two-body scattering  states as $m_1$ and $m_2$. Thus, the distribution function is given by
\begin{align}\label{dLF}
d_{LF}(s)   &=\frac{\pi \sigma(s)}{(s-m^2-\mathrm{Re}\Pi(s))^2+\mathrm{Im}\Pi(s)^2},
\end{align}
where $\mathrm{Re}\Pi(s)$ and $\mathrm{Im}\Pi(s)$ represent the real and imaginary parts of  $\Pi(s)$ respectively.
The spectral density function $\sigma(s)$ could be parameterized in various ways depending on the assumptions regarding the scattering amplitude. Here, we suggest a specific type of parameterization, in which the spectral density function $\sigma(s)$ is parameterized as
\begin{align}
{\label{eq:sigma1}}\sigma(s) &=|f(s)|^2=g^2\frac{E_1E_2}{\sqrt{s}}k^{2l+1}e^{-\frac{k^2}{k_0^2}}\theta(s-s_{th}).
\end{align}
The coupling vertex function  of the discrete state and the scattering channel, denoted as $f(s)$, is assumed to be real, where $g$ signifies the relative coupling strength. The energies of two final particles in the channel, $E_1$ and $E_2$, are expressed as $E_1=(s+m_1^2-m_2^2)/2\sqrt{s}$ and $E_2=(s-m_1^2+m_2^2)/2\sqrt{s}$ respectively. $m_1$ and $m_2$ are the respective masses of the two final particles. The relative momentum for particles in the center-of-mass frame of the scattering two-body channel, denoted as $k$, is expressed as $k=\sqrt{(s-(m_1+m_2)^2)(s-(m_1-m_2)^2)/4s}$. The symbol $l$ represents the relative angular momentum of the scattering two-body system. The unit step function, $\theta(s-s_{th})$, ensures the fulfillment of physical conditions.

The motivation for parameterizing the spectral density function in this manner is based on the following general considerations:
\begin{enumerate}
	\item It has been demonstrated that within the relativistic LF model, the self-energy function should be proportional to the relativistic phase space factor \cite{zhou:2020}
\bqa
k^2\frac{\mathrm d k}{\mathrm d E}=\frac{E_1E_2 k}{E},
\eqa
and the remaining portion should be proportional to $k^{2l}$ to ensure correct threshold behavior of the $T$-matrix. {{ It is worth emphasizing that the factor is consistently derived from the relativistic formalism of the LF model and the $s$-dependence of $E_1E_2$ factor plays an important role in this parameterization and make it applicable in both the equal-mass and unequal-mass two-body systems. This is also one of the distinct features of this parametrization from other parameterizations which empirically choose a factor of $k^{2l+1}/\sqrt{s}$ or just $k/\sqrt{s}$, which will usually fail in describing the unequal-mass system such as $\pi K$ or $\pi N$ systems.~\cite{Tornqvist:1995kr,Giacosa:2021mbz}  }}
	\item An exponential form factor $e^{-{k^2}/{k_0^2}}$, with $k_0$ being a suppression energy scale, is utilized to suppress high energy behavior. 		Rather than resorting to an empirical and artificial  cutoff factor as an upper integration limit, we can extend the dispersion integral $\Pi(s)$ to infinity aided by the exponential suppression, thereby ensuring convergence. Furthermore, the reasonability of using this exponential form factor finds its support in other contexts. For instance, in scenarios where the quark pair creation model is employed and the wave function of a quark potential model is utilized to describe meson states, the exponential factor naturally emerges in the coupling vertex with a more intricate structure.  From this perspective, the parameter $g$ can be regarded as representing the coupling strength and $1/k_0$ could be a parameter associated with the effective interaction range of the interacting mesons. {{Besides, the exponential factor might introduce singularities far away from the physical region, whose contribution to the physical observable is insignificant. Since we aim to present a parametrization method to represent the lineshape in the physical region and extract the nearby poles, this approximation is expected to be effective, as will be demonstrated through examples in subsequent sections.}}
\end{enumerate}

This parameterization has several fundamental properties. For a certain scattering channel, the resonance is entirely parameterized by three free parameters in a unitary $S$ matrix: the bare mass $m$, the coupling strength $g$, and the suppression energy scale $k_0$. For $s>s_{th}$, an obvious singular point arises at $s=s^{\prime}$ in the integrand of the dispersion integral. By applying the Cauchy integral theorem, the imaginary part can be readily deduced to be $\mathrm{Im}\Pi(s) = -\pi\sigma(s)$. The real part of the self-energy function corresponds to the principal value  of the dispersive integral, defined as $\mathrm{Re}\Pi(s)\equiv \bbint_{s_{th}}^{\infty}d s^{\prime}\frac{\sigma(s^{\prime})}{s-s^{\prime}}$, which is usually evaluated numerically.

The partial-wave scattering $S$-matrix is explicitly expressed as  	
\begin{equation}	
\mathcal S(s)=\frac{s-m^2-\mathrm{Re}\Pi(s)-i\pi\sigma(s)}{s-m^2-\mathrm{Re}\Pi(s)+i\pi\sigma(s)}.	
\end{equation}
This formulation evidently satisfies partial-wave unitary and can usually be parameterized as $S(s)=e^{2i\delta(s)}$, where $\delta(s)$ represents the phase shift. The inverse of the $\eta$ function could be regarded as the full propagator for the dressed state as
	\begin{equation}
	\mathcal{P}(s)=\eta^{-1}(s)=\frac{1}{s-m^2-\mathrm{Re}\Pi(s)+i\pi\sigma(s)}.
	\end{equation}
It is important to highlight that this form respects the analyticity of the general scattering amplitude. Upon analytically continuing the $S$ matrix to the complex $s$-plane,  more than one pole exist on the unphysical Riemann sheet. The nearby poles can be associated with physical states, sometimes manifesting as  the two-pole structure. Further elaboration on the analytical continuation will be provided in detail later on, as it is intricately linked to the central theme of this paper.

\emph{Non-Relativistic scenario}

For the non-relativistic scenario, the parameterization method is similar, but the formulas are expressed as functions of energy $E$:
\begin{align}
\mathcal S(E)   &=1-\frac{2\pi i \sigma(E)}{\eta(E)},\\
\eta(E)&=E-m-\Pi(E),\\
\Pi(E) &=\int_{E_{th}}^{\infty}d E^{\prime}\frac{\sigma(E^{\prime})}{E-E^{\prime}},
\end{align}
where $E_{th}=m_1+m_2$ is the energy threshold. If the non-relativistic two-body phase space factor is chosen to be the leading order nonrelativistic approximation of the relativistic version as
\bqa
E_1=m_1+\frac{k^2}{2m_1},\ \ E_2=m_2+\frac{k^2}{2m_2},\ \ E=E_1+E_2,\ \
\eqa
 the parameterization of spectral density function gives
\begin{align}
{\label{eq:sigma3}}\sigma(E)&=g^2\frac{E_1E_2}{E}k^{2l+1}e^{-\frac{k^2}{k_0^2}}\theta(E-E_{th}),\\
{\label{eq:momentum2}} k(E)&=\sqrt{2\frac{m_1m_2}{m_1+m_2}(E-m_1-m_2)}.
\end{align}
which is slightly different from the relativistic scenario and the $S$-matrix reads
	\begin{align}
		S(E)&=\frac{E-m-\mathrm{Re}\Pi(E)-i\pi\sigma(E)}{E-m-\mathrm{Re}\Pi(E)+i\pi\sigma(E)}.
	\end{align}

There is a notable advantage of the non-relativistic form of the parameterization. In the relativistic context,  the principal value of dispersion integral can only be numerically integrated, consuming substantial computational resources during fitting and even more so when extracting the poles on the complex $s$ plane. However, the dispersive integral of non-relativistic form can be explicitly represented by incomplete $\Gamma$ functions, as
\begin{align}
\Pi(E)&=\frac{k_0^{2l+3} (k^2+2 m_1 m_2)}{8 E k^4 m_1^2 m_2^2}      (k_0^2 (2 l+3) (k^2(m_1^2+m_2^2-m_1m_2)+ 2 m_1^2 m_2^2)+4 k^2 m_1^2 m_2^2) \Gamma (l+\frac{3}{2})
\nonumber \\ &
-\frac{(k^2+2 m_1^2) (k^2+2 m_2^2) (-k^2){}^{l+\frac{1}{2}}}{16 E m_1 m_2}  (2 l+3) (2 l+5)  e^{-\frac{k^2}{k_0^2}}    \Gamma (l+\frac{3}{2}) \Gamma (-l-\frac{5}{2},-\frac{k^2}{k_0^2})
\nonumber \\ &
-\frac{(m_1-m_2){}^2  {(2m_1 m_2)}^{l+\frac{1}{2}}}{4E}  (2 l+3) (2 l+5)  e^{\frac{2 m_1 m_2}{k_0^2}}   \Gamma (l+\frac{3}{2}) \Gamma (-l-\frac{5}{2},\frac{2 m_1 m_2}{k_0^2})
\end{align}
where  the incomplete Gamma function $\Gamma[a,z]$ is defined as $\Gamma[a,z]=\int_{z}^{\infty}t^{a-1}e^{-t}dt$. This analytical representation maintains the correct analytical structure, and its computation in the complex $E$ plane is notably fast, leading to substantial savings in computational time during subsequent calculations.
	
\subsection{The $(1,2)$ case V.S.  the Flatt\'{e} Parameterization}

When there exists one discrete state coupled to two non-degenerate channels, the $\eta$ matrix has only one element.
The dressed propagator of the discrete state  is given by
\bqa\label{propagator12}
\mathcal P(s)=\frac{1}{\eta(s)}= \frac{1}{s-m^2-\Pi_1(s)-\Pi_2(s)}
\eqa
where $\Pi_i(s) =\int_{s_{th,i}}^{\infty}d s^{\prime}\frac{\sigma_i(s^{\prime})}{s-s^{\prime}}$ represents the self-energy function of the $i$-th channel. Here, $\sigma_i$ and $s_{th,i}$ correspond to the respective spectral density function and the threshold energy squared for the $i$-th channel. The $S$-matrix element of two different channels could be represented in a matrix form
\bqa
\mathcal S=\left ( \begin{array}{ll} S_{11} & S_{12}   \\ S_{21} & S_{22} \end{array}\right )=\left ( \begin{array}{ll} \frac{s-m^2-\Pi_1(s)-\Pi_2(s)-2i\pi\sigma_1(s)}{s-m^2-\Pi_1(s)-\Pi_2(s)}& \frac{-2i\pi\sqrt{\sigma_1\sigma_2}}{s-m^2-\Pi_1(s)-\Pi_2(s)}   \\ \frac{-2i\pi\sqrt{\sigma_1\sigma_2}}{s-m^2-\Pi_1(s)-\Pi_2(s)} & \frac{s-m^2-\Pi_1(s)-\Pi_2(s)-2i\pi\sigma_2(s)}{s-m^2-\Pi_1(s)-\Pi_2(s)} \end{array}\right ).
\eqa

When $s_{th,1}<s<s_{th,2}$, since the self-energy $\Pi_2(s)$ is real, and $\Pi_1(s)$ possesses an imaginary part $-i\pi\sigma_1(s)$,  the element $S_{11}$, being the scattering matrix element of ``channel 1 $\rightarrow$ channel 1" can be expressed as
\bqa
S_{11}=\frac{s-m^2-\mathrm{Re}\Pi_1(s)-\Pi_2(s)-i\pi\sigma_1(s)}{s-m^2-\mathrm{Re}\Pi_1(s)-\Pi_2(s)+i\pi\sigma_1(s)},
\eqa
which is evidently unitary and can be represented by a pure phase as $S_{11}=e^{2i\delta_{11}}$. When $s>s_{th,2}$, both $\Pi_1(s)$ and $\Pi_2(s)$ contribute an imaginary part, leading to
\bqa
S_{11}=\frac{s-m^2-\mathrm{Re}\Pi_1(s)-\mathrm{Re}\Pi_2(s)-i\pi\sigma_1(s)+i\pi\sigma_2(s)}{s-m^2-\mathrm{Re}\Pi_1(s)-\mathrm{Re}\Pi_2(s)+i\pi\sigma_1(s)+i\pi\sigma_2(s)},
\eqa
which  is no longer unitary and it should be represented by a product of a phase and another inelasticity parameter.

Above the second threshold,  the element $S_{22}$, being the scattering amplitude of ``channel 2 $\rightarrow$ channel 2", can be explicitly written down as
\bqa
S_{22}=\frac{s-m^2-\mathrm{Re}\Pi_1(s)-\mathrm{Re}\Pi_2(s)+i\pi\sigma_1(s)-i\pi\sigma_2(s)}{s-m^2-\mathrm{Re}\Pi_1(s)-\mathrm{Re}\Pi_2(s)+i\pi\sigma_1(s)+i\pi\sigma_2(s)}.
\eqa\label{twochannel}
In general, the $S$-matrix satisfies the general coupled channel form
\bqa
\mathcal S=\left ( \begin{array}{ll} \zeta e^{2i\delta_1} & i(1-\zeta^2)^{1/2} e^{i(\delta_1+\delta_2)}   \\i(1-\zeta^2)^{1/2} e^{i(\delta_1+\delta_2)} & \zeta e^{2i\delta_2}\end{array}\right ).
\eqa
where $\zeta$ and $\delta_i$ are real and the inelasticity function with $\zeta\leq 1$. Due to the coupling to the second scattering channel, the $(1,2)$ parametrization contributes a cusp effect to the line shape of the amplitude. Thus the distribution function of the first channel is
\bqa\label{dLF2}
d_{LF}=\frac{\pi\sigma_1(s)}{(s-m^2-\mathrm{Re}\Pi_1(s)-\mathrm{Re}\Pi_2(s))^2+(\pi\sigma_1(s)+\pi\sigma_2(s))^2}
\eqa

Since  $\sigma_i(s)$ is defined as
\bqa\label{sigma12}
\sigma_i(s)=g_i^2\frac{E_{i,1}E_{i,2}}{\sqrt{s}}k_i^{2l_i+1}e^{-\frac{k_i^2}{k_0^2}}\theta(s-s_{th,i})
\eqa
which is similar to eq.(\ref{eq:sigma1}) and involves a factor of momentum $k_i$, this parametrization reflects the the characteristic features of the traditional  Flatt\'{e} parameterization. In the vicinity of the higher threshold, the $S$-wave propagator for the Flatt\'{e} form is given by \begin{equation}\label{eq:p-flatte}
\mathcal P_{Flatt\acute{e}}=\left\{ \begin{array}{ll} \frac{1}{m^2-s-im(g_1k_1+g_2k_2)} & s>s_{th,2}   \\ \frac{1}{m^2-s-im(g_1k_1+ig_2\sqrt{-k_2^2})} & s<s_{th,2}  \end{array}\right.
\end{equation}
It is evident from the dressed propagator of the $(1,2)$ case in Equation (\ref{propagator12}) that the terms $\mathrm{Re}\Pi_1(s)$ and $\mathrm{Re}\Pi_2(s)$ introduce a smooth $s$-dependent contribution, which, when focusing on a limited energy range, can be absorbed into the redefinition of the mass. The dependence on $k_i^{2l_i+1}$ within the $\sigma_i$ function effectively captures the cusp effect reminiscent of the Flatt\'{e} parametrization. Consequently, the current parameterization exhibits the analogous properties to those of the Flatt\'{e} parametrization.

\subsection{The $(2,2)$ case}

For the more complicated cases with two scattering channels, labeled with $1$ and $2$, and two discrete state, labeled with $a$ and $b$, the dressed propagator $\mathcal P=\eta^{-1}$ is  a $2\times 2$ matrix function
\begin{equation}
\eta(s)=\left(\begin{array}{cc}
s-m_a^2-\Pi_{aa,1}(s)-\Pi_{aa,2}(s)  & -\Pi_{ab,1}(s)-\Pi_{ab,2}(s) \\
 -\Pi_{ab,1}(s)-\Pi_{ab,2}(s) & s-m_b^2-\Pi_{bb,1}(s)-\Pi_{bb,2}(s) \end{array} \right),
\end{equation}
where $\sigma_{ab,i}=g_{ai}g_{bi}\frac{E_{i1}E_{i2}}{\sqrt{s}}k_i^{2l_i+1}e^{-\frac{k_i^2}{k_0^2}}\theta(s-s_{th,i})$.
\begin{align}
\mathcal S=\left (
\begin{matrix} S_{11} & S_{12}   \\ S_{21} & S_{22} \end{matrix}\right )
=\left ( \begin{matrix} 1-2\pi
i\sum\limits_{\alpha,\beta=a,b}\sigma_{\alpha\alpha,1}^{1/2}[\eta^{-1}]_{\alpha\beta}\sigma_{\beta\beta,1}^{1/2}&
-2\pi
i\sum\limits_{\alpha,\beta=a,b}\sigma_{\alpha\alpha,1}^{1/2}[\eta^{-1}]_{\alpha\beta}\sigma_{\beta\beta,2}^{1/2}\\
-2\pi
i\sum\limits_{\alpha,\beta=a,b}\sigma_{\alpha\alpha,1}^{1/2}[\eta^{-1}]_{\alpha\beta}\sigma_{\beta\beta,2}^{1/2}
& 1-2\pi
i\sum\limits_{\alpha,\beta=a,b}\sigma_{\alpha\alpha,2}^{1/2}[\eta^{-1}]_{\alpha\beta}\sigma_{\beta\beta,2}^{1/2}
\end{matrix}\right ).
\end{align}

When  several resonances are strongly coupled, there is no imperative need to segregate distribution functions akin to the single resonance scenario. Consider, for example,  processes as $e^+e^-\rightarrow Rs\rightarrow ``hadron\ \ pairs"$ involving several immediate $R$ resonances. Here,  the assumption is that the interaction vertex between $e^+e^-$ and each $R$ resonance is  weak, resulting in negligible contributions to the $\eta$ function. Consequently, the cross section for $e^+e^-\rightarrow\ \ ``channel\ \ i"$ is proportional to the scattering amplitude as
\bqa\label{crosssection}
[cross\ section](e^+e^-\rightarrow ``channel\ \ i")\propto |\sum_{\alpha,\beta}g_{e\alpha}[\eta^{-1}]_{\alpha\beta}f_{\beta i}|^2.
\eqa
in a manner of the perturbative method.

Extending this method to scenarios with multiple discrete states and multiple scattering channels is a straightforward process, which also maintains the fundamental properties of the  method in evaluating scattering amplitudes.

	\section{Analytic continuation and pole structure}\label{analyticcontinuation}

As previously discussed, the poles of $S$-matrix located in the unphysical sheet  reveal essential information about the nature of resonances rather than the lineshape itself. 
Each self-energy function, as depicted in eq. (\ref{eq:selfenergy}), introduces a discontinuity along the real axis extending from the associated threshold to infinity in the $S$ matrix. In cases involving $n$ non-degenerate decaying channels, there exist $n$ branch cuts and  $2^n$ Riemann sheets. However, it is crucial to note that only the poles situated on the Riemann sheet nearest to the physical region significantly influence the lineshape of the resonance. These poles on the closest Riemann sheet are the primary contributors to the physical behavior of the system.

\subsection{Two-pole structure in the $(1,1)$ case}

In the case that one discrete state couples to one decaying channel, there is only one unitarity cut which divides the complex $s$-plane to two Riemann sheets. The $\eta$ function on the first Riemann sheet~(phyical sheet) is defined as
\bqa
\eta^{I}(s)=s-m_0^2-\int_{s_{th}}^{\infty}d s^{\prime}\frac{\sigma(s^{\prime})}{s-s^{\prime}},
\eqa
while that on the second Riemann sheet is
\bqa
\eta^{II}(s)=s-m_0^2-\int_{s_{th}}^{\infty}d s^{\prime}\frac{\sigma(s^{\prime})}{s-s^{\prime}}-2\pi i\sigma(s).
\eqa
The poles of the $S$-matrix element on the second Riemann sheet can, in principle, be determined by solving
\bqa
\eta^{II}(s_0)=0,
\eqa
where $\sqrt{s_0}=M_{pole}-i\Gamma_{pole}/{2}$ with $M_{pole}$ and $\Gamma_{pole}$ representing the pole mass and pole width.  Notably,  the schwartz reflection property $\mathcal{S}(s^*)=\mathcal{S}^*(s)$ requires the existence of  another conjugate pole $s^*_0$ on the upper-half complex plane.  For clarity and to avoid unnecessary complexity, the discussion will focus on the primary pole without delving into the details of its conjugate counterpart unless explicitly required for a specific analysis.

We reiterate a crucial point that, contrary to conventional beliefs, the dressed propagator $\mathcal{P}(s)=\eta^{-1}(s)$ may exhibit multiple resonance poles, as evidenced by various illustrative examples presented in references \cite{Xiao:2016dsx, Xiao:2016mon, Xiao:2023lpv}. Among these poles, one emerges from the bare state in the conventional manner, acquiring its width through interactions with scattering channels. However, an additional dynamically generated pole, distinct from the conjugate of $s_0$, is also induced by the coupling effects.
When the coupling is sufficiently strong, the pole manifests observable effects and can be typically parameterized as an isolated resonance using the traditional Breit-Wigner formalism in experimental analyses. Conversely, under weak coupling, this dynamically generated pole might contribute mildly, and is often assimilated into the background component.
In some special example, the dynamically generated one is a very broad resonance and supplies an unconventional contribution to the phase shift or cross section data  which is hard to be discriminated from the background. As a result, there are  controversies on its existence or nature.  Notably, examples like  $f_0(500)$ and $K_0^*(700)$  fall into this category, which have generated extensive discussions in numerical analyses to elucidate their nature further.

Furthermore, we propose that this parametrization derived from the extended Lee-Friedrichs model effectively captures the essence of two-pole structures and allows for data parametrization using fewer variables compared to the conventional Breit-Wigner approach. Unlike the mass and width parameters in the Breit-Wigner formalism, which lack intrinsic physical significance, the mass parameter $m$ in this context can be interpreted as the bare mass of the state in the ``free" Hamiltonian when the interaction between the bare state and the coupled channel is turned off. Thus, it holds the potential to be compared against predictions from quenched quark models, providing a meaningful connection to theoretical frameworks beyond mere empirical fitting.

\subsection{The phase shift and line shape of the two-pole structure}

The scattering amplitude in  the $(1,1)$ case with the real $s$ can be expressed as
\begin{equation}	
\mathcal S(s)=\frac{\eta^{II}}{\eta^{I}}=\frac{m^2-\bbint_{s_{th}}^{\infty}d s^{\prime}\frac{\sigma(s^{\prime})}{s^{\prime}-s}-s+i\pi\sigma(s)}{m^2-\bbint_{s_{th}}^{\infty}d s^{\prime}\frac{\sigma(s^{\prime})}{s^{\prime}-s}-s-i\pi\sigma(s)},	
\end{equation}
which is evidently unimodular and can be represented by a pure phase $S(s)=e^{2i\delta(s)}$. Typically, the phase $\delta(s)$ exhibits asymptotic behavior such that $\delta(s_{th})=0$ and $\delta(\infty)\rightarrow\pi$. However, there often exists a conventional pole along with another dynamically generated pole~(two-pole structure) on the complex $s$ plane.

\begin{figure}[htbp]
	\centering
	\includegraphics[width=5cm]{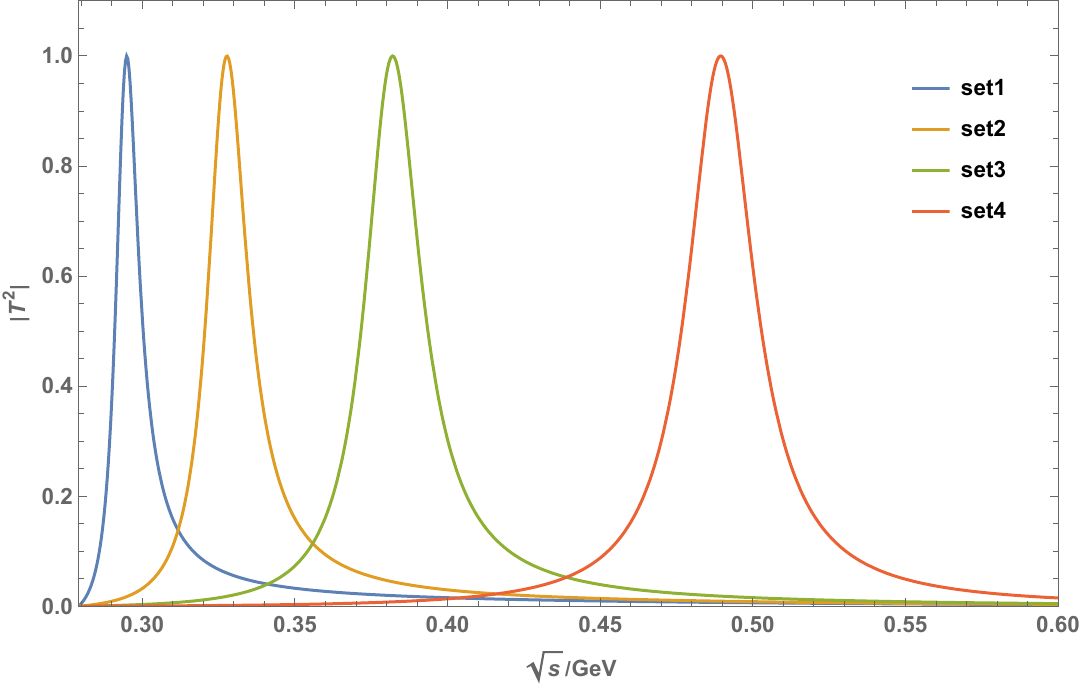}\put(-130,80){\bf (a)}
	\includegraphics[width=5cm]{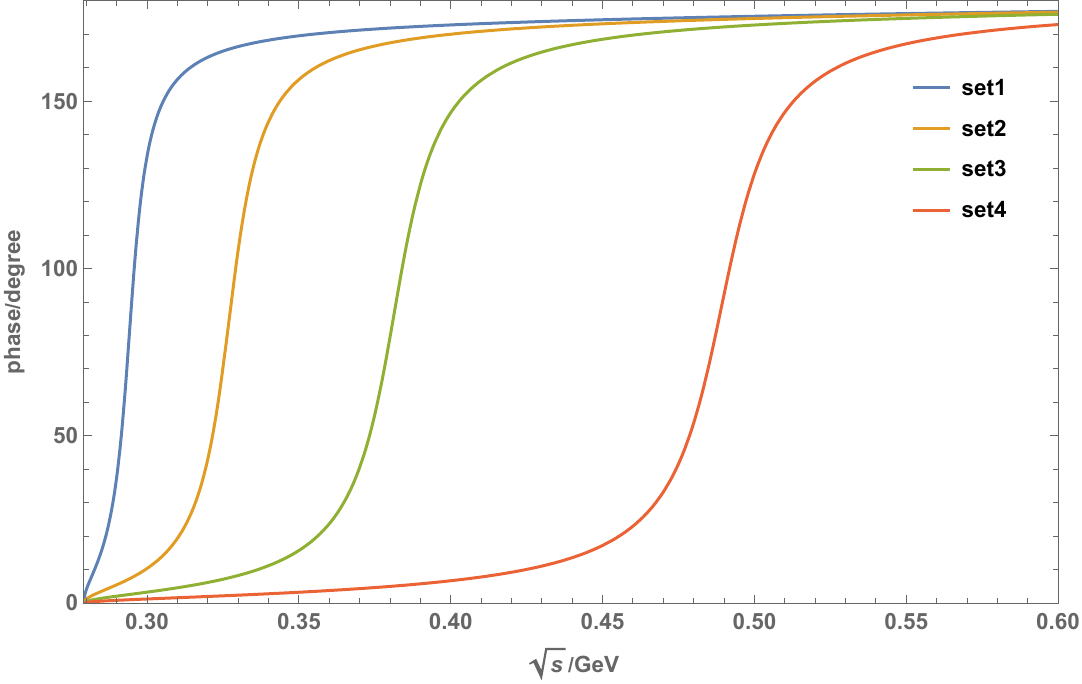}\put(-130,80){\bf (b)}
	\caption{$|T|^2$ and phase for partial $S$-wave scattering with $m=0.32, 0.35, 0.4, 0.5$, respectively, $k_0=0.3$ and $g=0.5$.}
	\label{shapev2}
\end{figure}

\begin{figure}[htbp]
	\centering
	\includegraphics[width=5cm]{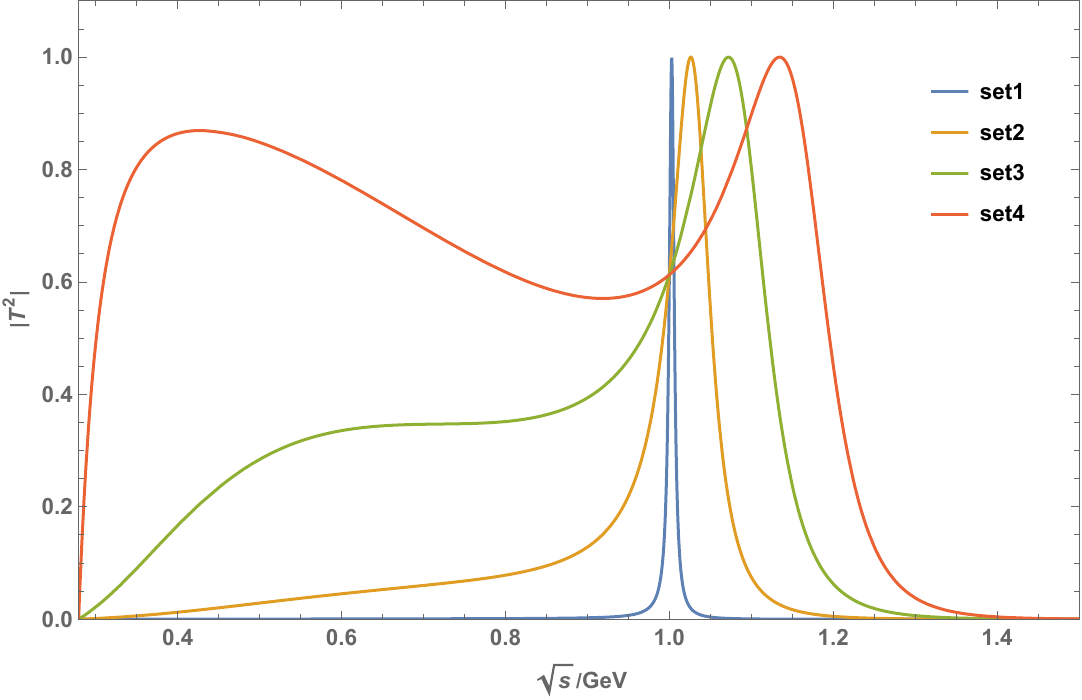}\put(-130,80){\bf (a)}
	\includegraphics[width=5cm]{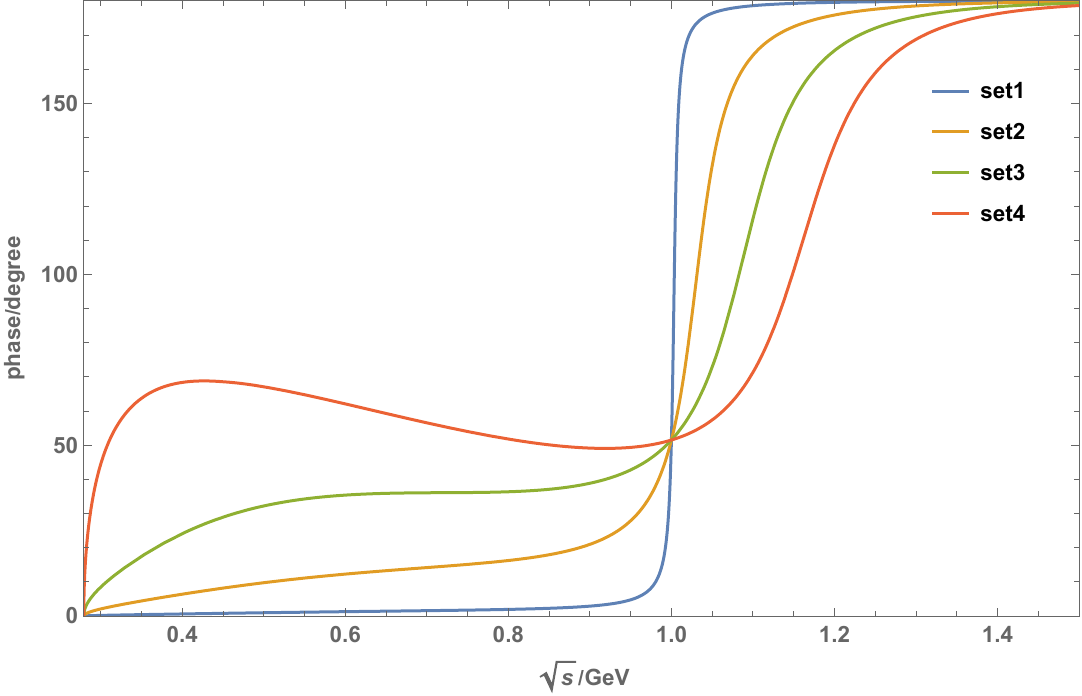}\put(-130,80){\bf (b)}
	\caption{$|T|^2$ and phase for partial $S$-wave scattering with $m=1.0$, $k_0=0.3$ and $g=0.5, 1.5, 2.5, 3.5$ respectively.}
	\label{shapev1}
\end{figure}

We illustrate several exemplary figures depicting the typical line shapes and phase shifts of partial $S$-wave scattering amplitude  with different sets of parameter values in Figure~\ref{shapev2} and \ref{shapev1}. Here,  $T=(S-1)/2i$  and $|T|^2$ can be easily connected with experimental observations of cross section or events.
Typically, when the coupling is weak and the bare state is distant from the threshold, the line shape appears narrow and symmetric, akin to the traditional Breit-Wigner line shape, as shown in part (a) of Figure~\ref{shapev2}. The nearby pole contributes a sharp rise in the phase shift, as shown in (b) of Figure~\ref{shapev2}. However, as the coupling becomes stronger  or the bare state is close to the threshold, this lineshape is asymmetric and the phase shift deviates from the Breit-Wigner form, exemplified in part (a)  and (b) of Figure~\ref{shapev1}. In some circumstances,  the dynamically generated pole may approach the physical region, giving rise to an additional resonance peak in the $|T|^2$ line shape. Correspondingly, the  phase shift undergoes significant distortion. In the following numerical analysis section,  diverse examples will be presented to exhibit these distinct behaviors.
	
\subsection{The poles of the case with  multiple scattering channels}

When considering the importance of the coupled channel effects in the parameterization, every scattering channel contributes a discontinuity of the self-energy function in eq.~(\ref{eq:selfenergy}). Consequently, every element of the  $\eta$ function matrix possesses identical branch cuts to those of the $S$ matrix elements.
 With each cut doubling the number
of Riemann sheets, an amplitude with $n$ thresholds will manifest $2^n$ Riemann sheets. Nonetheless, only the poles on the nearest Riemann sheets to the physical region significantly influence  the experimental observations.
In this study, we
denote the Riemann sheet connected to the physical region between the $i$-th
and the $(i+1)$-th threshold as the $(i+1)$-th sheet. The last relevant sheet pertains to the physical region above the highest threshold. The $\Pi_{ab}$ function in eq.~(\ref{eq:eta}) on the $(i+1)$-th sheet is generally expressed as
\bqa
\Pi_{ab}^{(i+1)}(s)=\Pi_{ab}(s)+\sum_{n=1}^{i}2\pi i\sigma_{ab,n}(s).
\eqa
The other unphysical sheets are distant from the physical
region, making poles on these sheets inconsequential to the analysis and thus not under consideration.

Consequently, given the matrix form of the scattering amplitude, the poles of the scattering amplitude on the $N$-th Riemann sheet correspond to the zeros of the determinant of the $\eta$ matrix on the same sheet. Therefore, the pole positions can be determined by solving the equation
\bqa\label{deteta}
\mathrm{Det}[\eta^{(N)}(z_0)]=0,
\eqa
on the complex energy plane, where $z_0=(M-\frac{\Gamma}{2}i)^2$ where $M$ and $\Gamma$ represent the pole mass and width respectively.

\section{Numerical examples}\label{numerical}

In experiment analysis, the scattering phase shift contributed by the resonance is represented by the simplest partial wave $S$ matrix element as $\delta(s)=\mathrm{arg}[S(s)]/2$. This can also be expressed as $\delta(s)=\mathrm{arg}[\eta^{II}(s)]=-\mathrm{arg}[\eta^{I}(s)]$ in the elastic region. The cross section data and mass distribution data are proportional to the distribution function of resonances, as presented in eq.~(\ref{crosssection}).  In scenarios where there is only one discrete bare state present, the cross section is typically proportional to the distribution function $d_{LF}$ as outlined in eqs.(\ref{dLF}) or (\ref{dLF2}), alongside other background contributions.

We will show several typical examples, including the $\rho$ meson, $\Delta(1232)$ baryon, $K_0^*$ mesons, and $f_0$ mesons, each characterized by distinct production processes. During data fitting, background contributions are omitted unless deemed necessary. The goodness of fit is quantified by the expression
\bqa
\chi^2/dof=\frac{1}{N_{exp}-N_{par}}\sum(\frac{y_{exp}(E_i)-y_{th}(E_i)}{\Delta y_{exp}})^2,
\eqa
where $y_{exp}$, $\Delta y_{exp}$ and $y_{th}$ represent the experiment value, its uncertainty and the theoretical value of data point, respectively. Here,  $N_{exp}$ and $N_{par}$ denote the total number of  data points and the number of parameters in the fit.

It is crucial to emphasize that our intention is not to conduct a comprehensive experimental analysis of specific processes but rather to employ them as illustrative examples of the current parameterization approach. For a deeper understanding of the resonances' nature, a more detailed and elaborate analysis is essential.

We have chosen not to discuss the shortcomings of the traditional BW parametrization with a constant $\Gamma$, as this has been extensively addressed in the literature. In this paper, our focus is on comparing the outcomes of the improved BW parametrization form incorporating the Blatt-Weisskopf factor, commonly employed in experimental analyses, with those of the extended Lee-Friedrichs parametrization. It is essential to note that beyond the standard mass parameter $M_{BW}$ and width parameter $\Gamma_{BW}$, an additional scale parameter $q_0$ is introduced for the  improved BW parameterization with energy dependence in high partial wave, as discussed~\cite{ParticleDataGroup:2022pth}.

\subsection{$\rho(770)$ \label{rhoSection}}

The $\rho(770)$ contributes a typical resonance shape to the phase shift data of $\pi\pi$ $IJ=11$ in $\pi^+p\to\pi^+\pi^-\Delta^{++}$ and $\pi^+p\to K^+K^-\Delta^{++}$ reactions~\cite{Protopopescu:1973sh} or the distribution of $\pi\pi$ in the  $\tau\rightarrow \pi\pi_0\nu_\tau$  process~\cite{Davier:2013sfa}. Since these data sets are acquired through different procedures, we fit them individually.

\begin{figure}[htbp]
	\centering
\includegraphics[width=7cm]{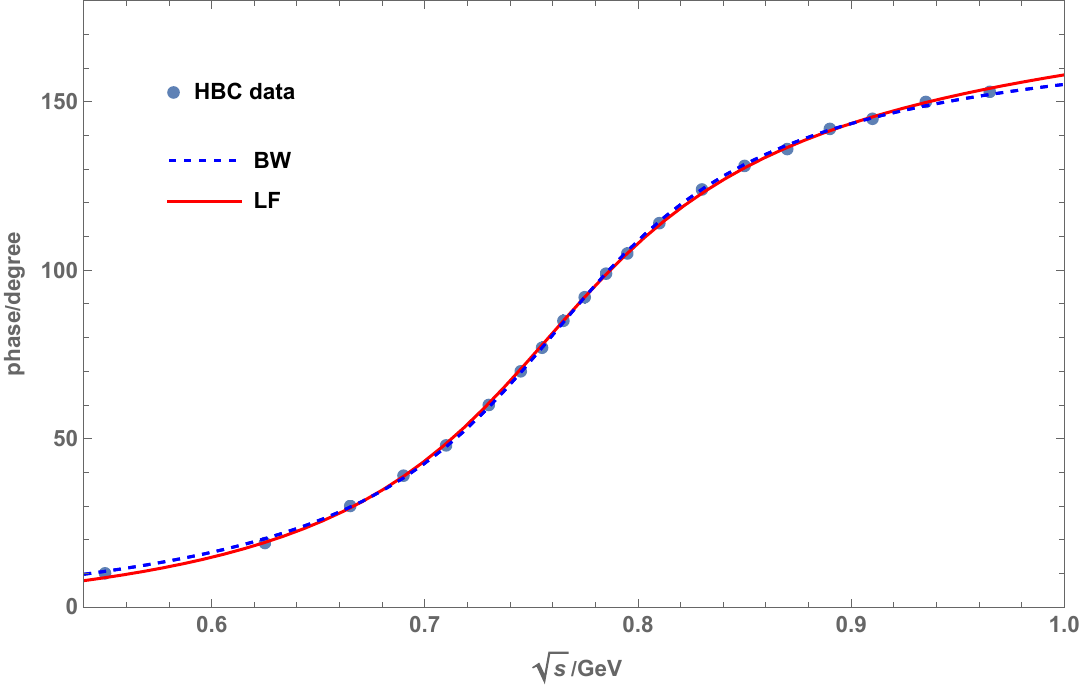}
\includegraphics[width=5cm]{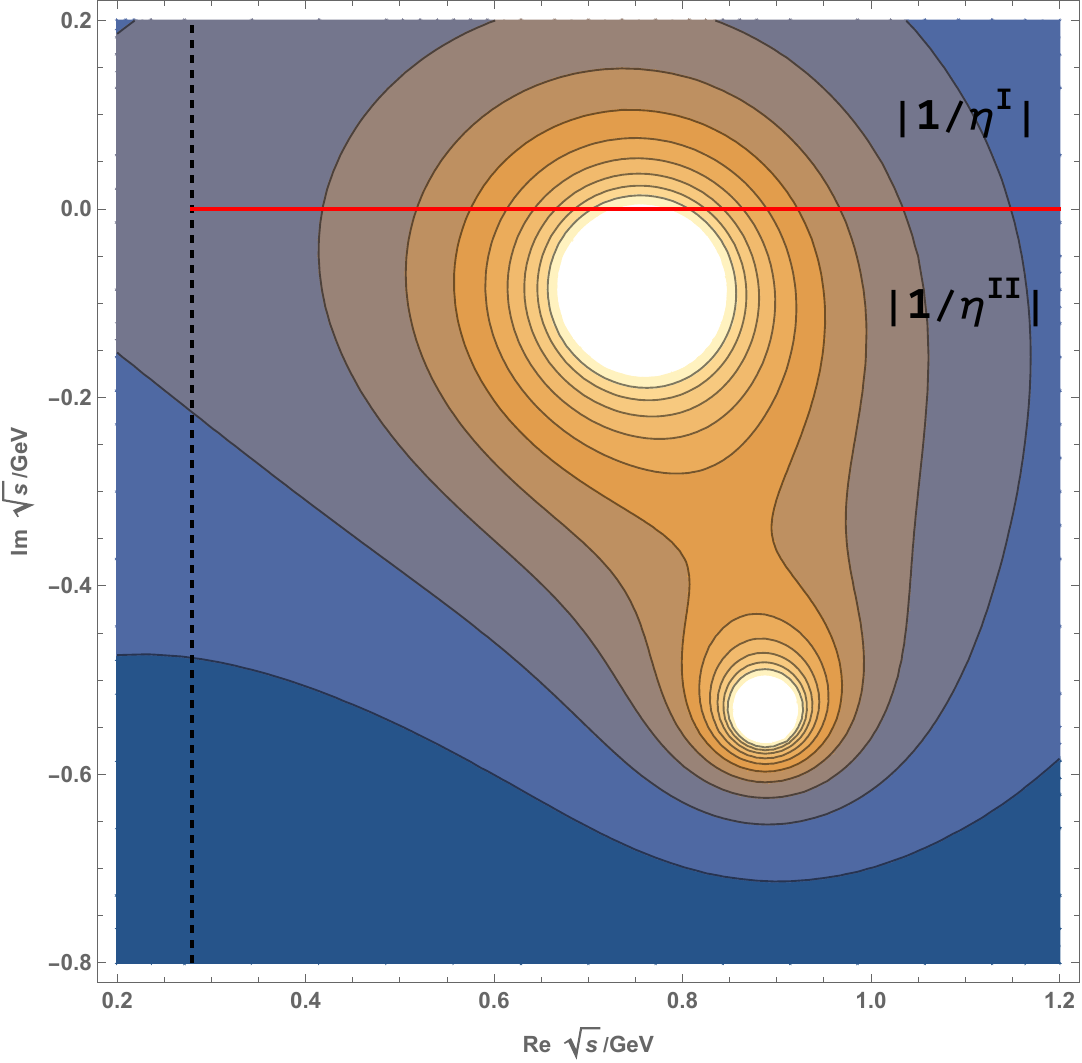}

	\caption{The left panel is the $IJ=11$ $\pi\pi$ scattering phase shift data~\cite{Protopopescu:1973sh} and the fit results  according to the LF parameterization and the BW parameterization. The right panel is the contour plot of $|1/\eta(\sqrt{s})|$ in the LF form.}
	\label{fig-fit-rho}
\end{figure}

For the fit of the $IJ=11$ $\pi\pi$ phase shift $\delta_1^1$~\cite{Protopopescu:1973sh}, we utilize the parameterization of $(1,1)$ case to describe the data with three free parameters: $m$, $k_0$ and $g$.  The fit outcomes and the extracted pole positions are shown in Figure~\ref{fig-fit-rho} and Table~\ref{tab-add-rho}, with the mean values and uncertainties of fitted parameters assessed by the Minuit2 programme. The uncertainties for the fitted parameters are in parentheses, for example, 0.829(10) means 0.829$\pm$0.010.

The fit result is in good agreement with the data point with $\chi^2/dof<0.5$ even though the uncertainties of the data is very small.
Besides the pole at $\sqrt{s_{p1}}=757^{+50}_{-65}+\frac{i}{2} 164_{-20}^{+28}$, which is consistent with the resonance pole representing $\rho(770)$ in BW parameterization, another broad pole located at $\sqrt{s_{p2}}=888^{+117}_{-102}+\frac{i}{2} 1074_{-86}^{+102}$ is also found in the complex $s$ plane as shown by the contour plot $|1/\eta|$ in complex $\sqrt{s}$ plane. The broad one is located notably far away from the real axis and contributes slightly to physical observables, whose role will be discussed later.

\begin{table}[htbp!]
	\caption{Fit results for $\rho(770)$ resonance from $\pi^+\pi^-$ phase shift $\delta_1^1$ data.  }
	\vspace{0.2cm}
	\label{tab-add-rho}
	\centering
	\footnotesize
	\begin{tabular}{c|cccccc|ccc}
		\hline\hline
		Fit  & $\chi^2/dof$ & $m$ & $k_0$ & $g$ &   &  & $\sqrt{s_{R,1}^{II}}$ & $\sqrt{s_{R,1}^{II}}$ & $M_{BW}-i\Gamma_{BW}/2$\\	
		\hline
phase shift: FL	    & 0.49 & 0.829(10) & 0.34(2) & 3.45(23) &   &   & $757^{+50}_{-65}-i82^{+14}_{-20}$ & $888^{+117}_{-102}-i537^{+51}_{-43}$ & -\\
phase shift: BW     & 0.51 &  0.772(3) & 0.66(27) & 1.82(19) &  &   & - & - &   $(772\pm3)-i(77\pm22)$\\	
		\hline\hline
	\end{tabular}
\end{table}

\begin{figure}[htbp]
	\centering
\includegraphics[width=7cm]{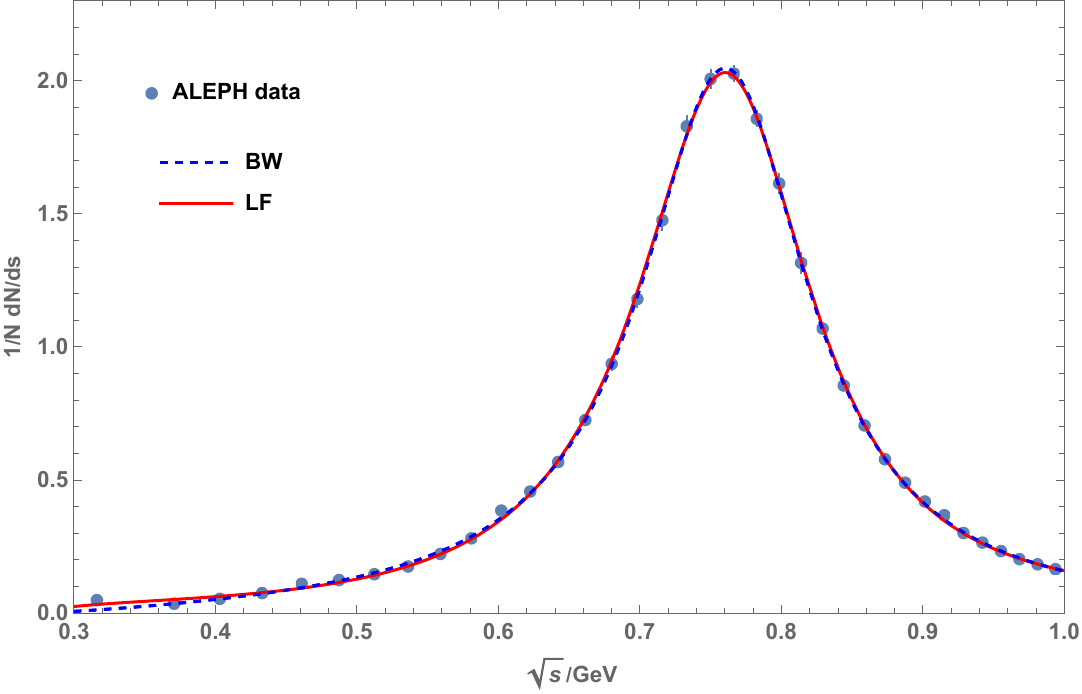}	
\includegraphics[width=5cm]{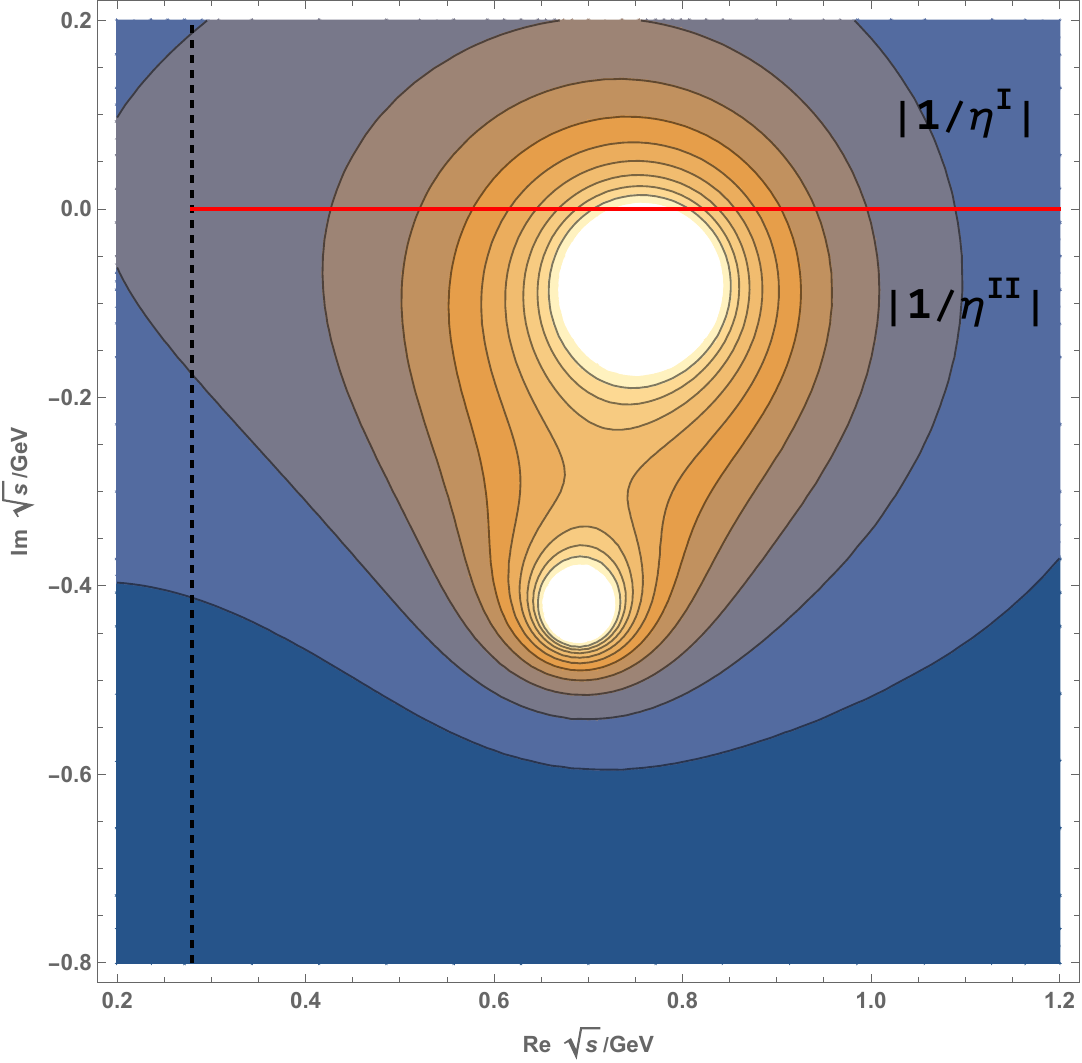}
    \caption{Left: the $\tau$ spectral function data together with the fit results of $\rho(770)$ resonance in the LF form and the BW form. Right: the contour plot of $|1/\eta|$ in the LF parametrization.}
	\label{fig-fit-rho-dist}
\end{figure}

The $\tau$ spectral function in the  $\tau\rightarrow \pi\pi^0\nu_\tau$  process, as documented in~\cite{Davier:2013sfa}, is also subjected to fitting in the LF parameterization, with  an additional simple smooth background contribution proportional to the kinematic factor assumed.  Thus, the spectral function is parameterized as $\mathcal{F}=c_1 d_{LF}+c_2 k/E$, and the fit result is depicted in Figure~\ref{fig-fit-rho-dist}, with the parameters and poles' positions obtained listed in Table~\ref{tab-res-rho-dist}. In this fit, the narrow pole aligns closely with the narrow one identified in the fit of phase shift data, while the position of the broader one is a little different. {{Stableness of the narrow pole suggests that the LF parametrization adeptly captures the poles situated near the real axis in scattering or production amplitudes. The broad one, contributing marginally to observables in the physical region, exhibit a degree of sensitivity to the modeling of the background contribution.

Despite
the common belief that $\rho(770)$ resonance necessitates a single pole representation, we refrain from labeling the additional pole as spurious.
In our view point, the quantum field theory leaves room for the existence of extra dynamically generated poles in the complex energy plane of the scattering amplitudes.
On the experimental side, the choice of how many poles to include  depends on the need to accurately represent the data while avoiding unnecessary complexity introduced by additional Breit-Wigner parameters.

Furthermore, in subsequent instances such as $\Delta(1232)$ and $K_0^*$, it becomes evident that the secondary pole  plays an important role in data description, suggesting that it could be  recognized as a non-spurious entity.
From a practical perspective, as this parametrization caters to experimental analysis, one may  ignore the poles with $\mathrm{Im}E>800$MeV, base on subjective assessment of their relevance. Nevertheless, the presence of such poles presents an intriguing prospect for future theoretical investigations.}}

\begin{table}[htbp!]
	\caption{Fit results for $\rho(770)$ resonance from $\pi\pi$ distribution data.  }
	\vspace{0.2cm}
	\label{tab-res-rho-dist}
	\centering
	\footnotesize
	\begin{tabular}{c|cccccc|ccc}
		\hline\hline
		Fit  & $\chi^2/dof$ & $m$ & $k_0$ & $g$ & $c_1$ & $c_2$& $\sqrt{s_{R,1}^{II}}$(MeV) & $\sqrt{s_{R,1}^{II}}$(MeV) & $M_{BW}-i\Gamma_{BW}/2$(MeV)\\	
\hline
distribution: FL    & 0.69 & 0.784(5) & 0.27(1) & 4.18(29) & 0.056(2) & 0.13(4) & $762^{+20}_{-40}-i76^{+32}_{-24}$ & $689^{+72}_{-53}-i430^{+36}_{-45}$ & -\\
distribution: BW    & 0.54 &  0.763(3) & 0.21(13) & 3.10(76) & 0.22(1) & 0.015(53) & - & - & $(763\pm3)-i(72\pm79)$\\	
		\hline\hline
	\end{tabular}
\end{table}

For comparison, we additionally provide the fit outcome of the improved BW parameterization incorporating an energy-dependent Blatt-Weisskopf barrier factor, as defined in Eq.~(\ref{improvedBW}). It is observed that the fit qualities of both the LF form and the improved BW form are comparable, utilizing the same number of parameters. The $k_0$ parameter for the BW form in Table~\ref{tab-res-rho-dist} represents the scale parameter.

\subsection{ $\Delta(1232)$ \label{DeltaSection}}

The $\Delta(1232)$ is recognized as the lowest $I=3/2$ baryon resonance state, displaying an asymmetric lineshape in the cross-section data in the $P_{33}$ wave of $\pi N$ scattering, as documented in~\cite{Carter:1971tj}. The phase shift data associated with this resonance also deviate from the typical behavior expected for a resonance. Conventionally, the data is attributed to the $\Delta(1232)$ resonance, characterized by a BW mass of 1232 MeV and a width of about 117 MeV, aligning with the peak of the cross-section, alongside a smoothly varying background.
However, various partial wave analyses reveal a pole position situated around $1210 - \frac{i}{2}100$ MeV~\cite{Svarc:2014zja,Anisovich:2011fc,Cutkosky:1980rh}, which is also supported by a recent study based on the Roy-Steiner equation\cite{Hoferichter:2023mgy}. This result is different from the conventional Breit-Wigner parameters, which suggests a more intricate resonance structure for the $\Delta(1232)$ state within the $\pi N$ system.

\begin{figure}[htbp]
	\centering
\includegraphics[width=5cm]{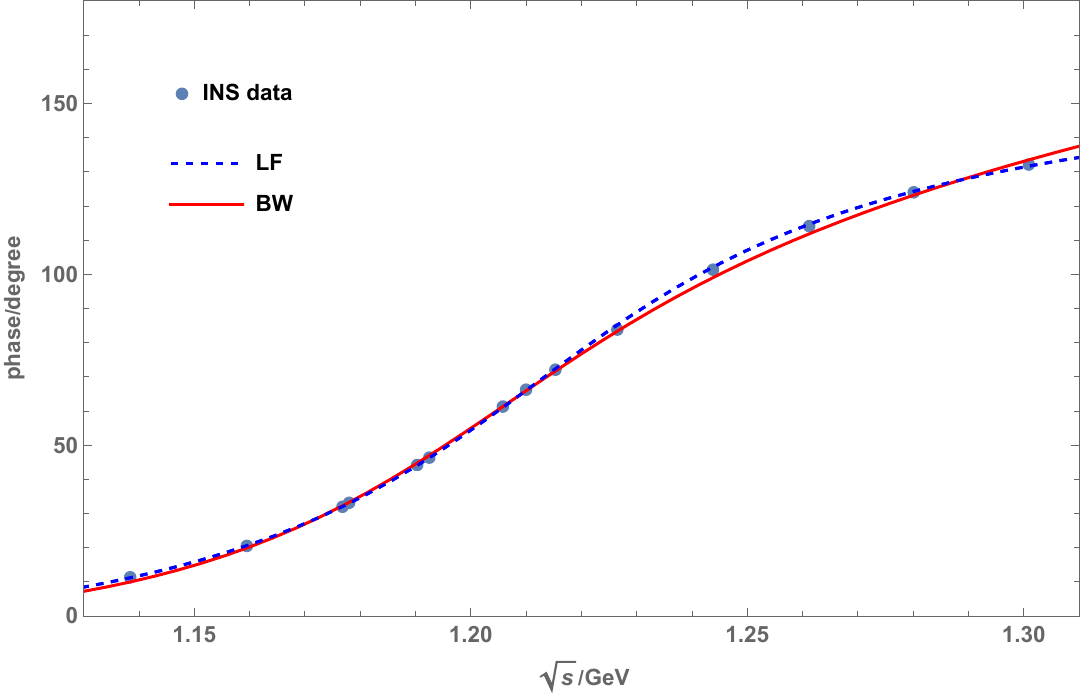}
\includegraphics[width=5cm]{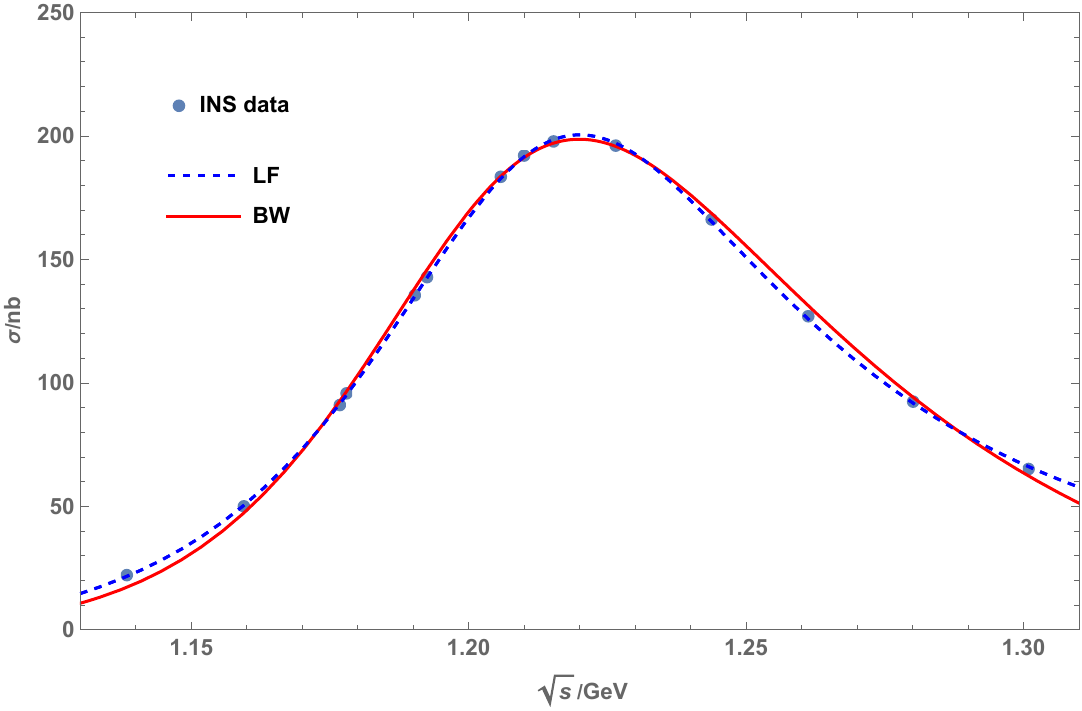}
\includegraphics[width=3.4cm]{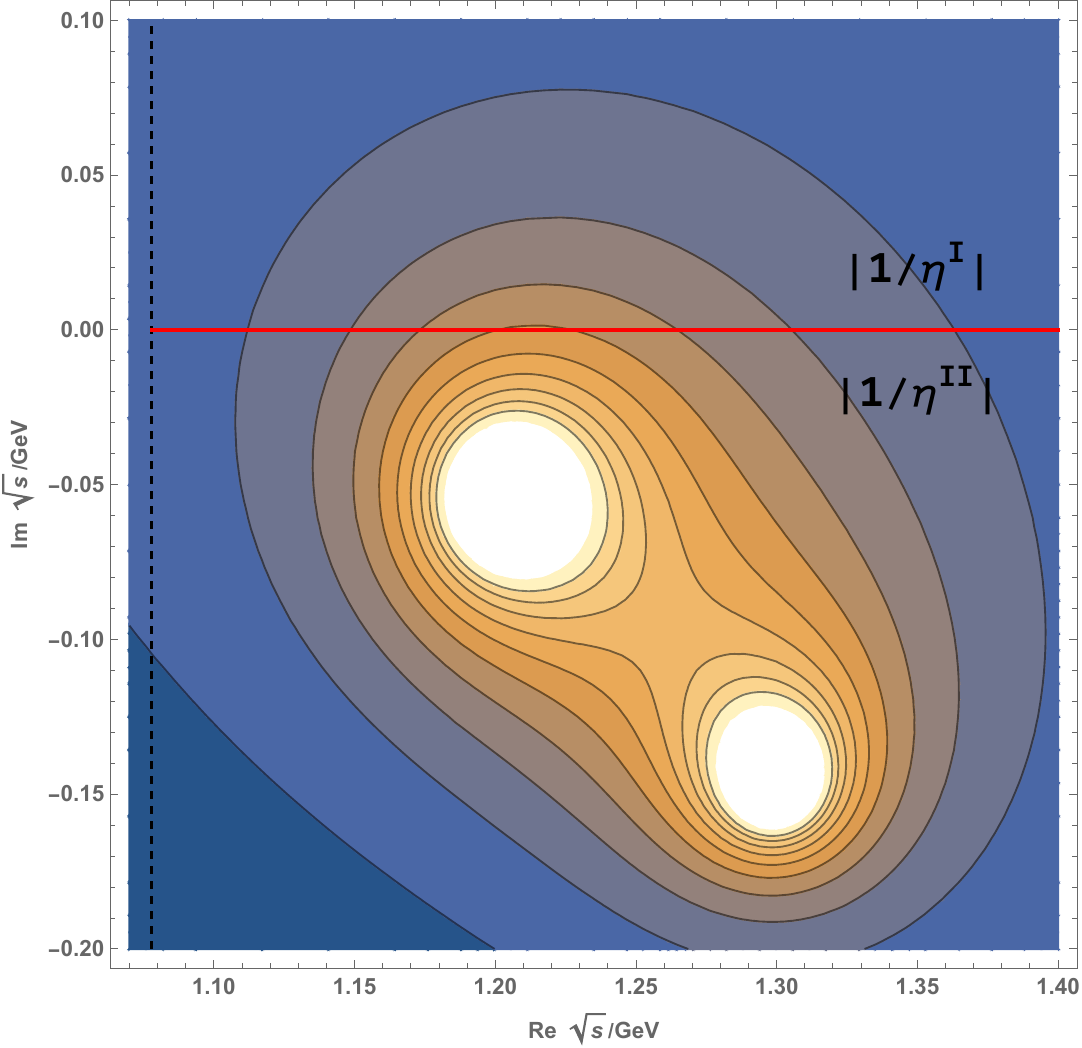}
	\caption{The phase shift~(left) and  cross section data~(middle) in $\pi p$ scattering~\cite{Carter:1971tj}, and the contour plot of $|1/\eta|$(right) in the LF parametrization.}
	\label{fig-fit-delta}
\end{figure}

In present study,  a simultaneous fit for the phase shift and cross section data of  $\pi N$ scattering data, as depicted in ref.~\cite{Carter:1971tj},  was performed using the LF form with the BW form for  comparison, without presuming the background contributions.  The cross section was parameterized as $\sigma=\frac{8\pi\hbar^2}{k^2}|T^2|$, where $k$ denotes the relative momentum in relativistic scenario. In this case, $\chi^2/dof$ of the fit with LF form is larger than that of the BW form, a discrepancy attributed in part to the high precision of the data set. {{in comparison with the fit to $\Delta(1232)$ using the Sill distribution as presented in ref.~\cite{Giacosa:2021mbz}, the LF parametrization is capable of describing the data with greater accuracy.}} Beside this, the line shape portrayed by the LF parameterization  remains consistent with the data set, as illustrated in Figure \ref{fig-fit-delta}. The fit results, based solely on three parameters, $m_0$, $k_0$ and $g$, are listed in Table~\ref{tab-res-delta}, along with the pole position extracted from the $\eta$ function.  It is crucial to highlight that the position of the first pole, located at $(1204\pm4)-\frac{i}{2}(110\pm4)$MeV, is consistent with those derived from other partial wave analyses as refs. \cite{Svarc:2014zja,Anisovich:2011fc,Cutkosky:1980rh,Hoferichter:2023mgy}.

\begin{table}[htbp!]
	\caption{Fit results for $\Delta(1232)$ resonance from $\pi p$ phase shift and cross section data in ref.~\cite{Carter:1971tj} in FL parameterization. }
	\vspace{0.2cm}
	\label{tab-res-delta}
	\centering
	\footnotesize
	\begin{tabular}{c|cccc|ccc}
		\hline\hline
		Fit  & $\chi^2/dof$ & $m$ & $k_0$ & $g$ & $\sqrt{s_{R1}^{\Rmnum{2}}}$ (MeV) & $\sqrt{s_{R2}^{\Rmnum{2}}}$ (MeV) &$M_{BW}-i\Gamma_{BW}/2$(MeV)  \\
		\hline
LF  & 36.4 & 1.2604(7) & 0.209(2) & 6.76(5) & $(1204\pm4)-i(55\pm2)$ & $(1300\pm7)-i(145\pm4)$ &- \\
BW  &  2.5 & 1.2310(4) & 0.215(7) & 4.97(9) & - & - & $(1231.0\pm0.4)-i(55\pm3)$ \\
		\hline\hline
	\end{tabular}
\end{table}

{{ Moreover, a broader resonance pole emerges around $(1300\pm7)-\frac{i}{2} (290\pm8)$MeV, as depicted in the contour plot. Notice  that there is no typical resonance line shape in the $P_{33}$ $\pi N$ scattering data around 1300$\sim$1600 MeV. In the literature, different multichannel analysis reveal that the $\pi N$ scattering data in $P_{33}$ wave required another $\Delta(1600)$ resonance, though different pole positions were obtained by different groups, with a pole mass of about $1470\sim 1590 $MeV and a pole width of $150\sim 320$MeV as quoted in the PDG table~\cite{ParticleDataGroup:2022pth}. For example,  the extended partial-wave analysis of $\pi N$ scattering by the George Washington group~(referred to as SP06 solution)  reported an additional pole at $1457-\frac{i}{2}400$MeV in the $P_{33}$ wave~\cite{Arndt:2006bf}, while another coupled-channel analysis including the $K\Sigma$ photoproduction report a pole at about $1590-\frac{i}{2}136$MeV~\cite{Ronchen:2022hqk}. Perhaps due to the large discrepancy among these results, this $\Delta(1600)$ state is classified as a three-star state in $\pi N$ channel in the PDG.

With this in mind, we also performed a fit to the real and imaginary part of $T$-matrix element of SP06 without assuming the background contribution. The data below the $\eta N$ threshold could be described well as shown in Fig.\ref{fig-fit-delta-WI08}. The lower pole is still located at about $1202-\frac{i}{2}88$MeV, while the higher one is located at about $(1370\pm4)-\frac{i}{2}(378\pm4)$MeV, with a larger mass and width than the previous result. Compared with the results of SP06, the broader pole has a lighter mass and similar width, which might be because this is only a simplified single-channel fit without assuming the background contribution.  A comprehensive analysis may also need to include the coupled channel effect. It is plausible that  $\Delta(1232)$ and   $\Delta(1600)$ might be a two-pole structure. The full analysis lies beyond the scope of this work and is deferred to future studies.
}

\begin{figure}[htbp]
	\centering
\includegraphics[width=5cm]{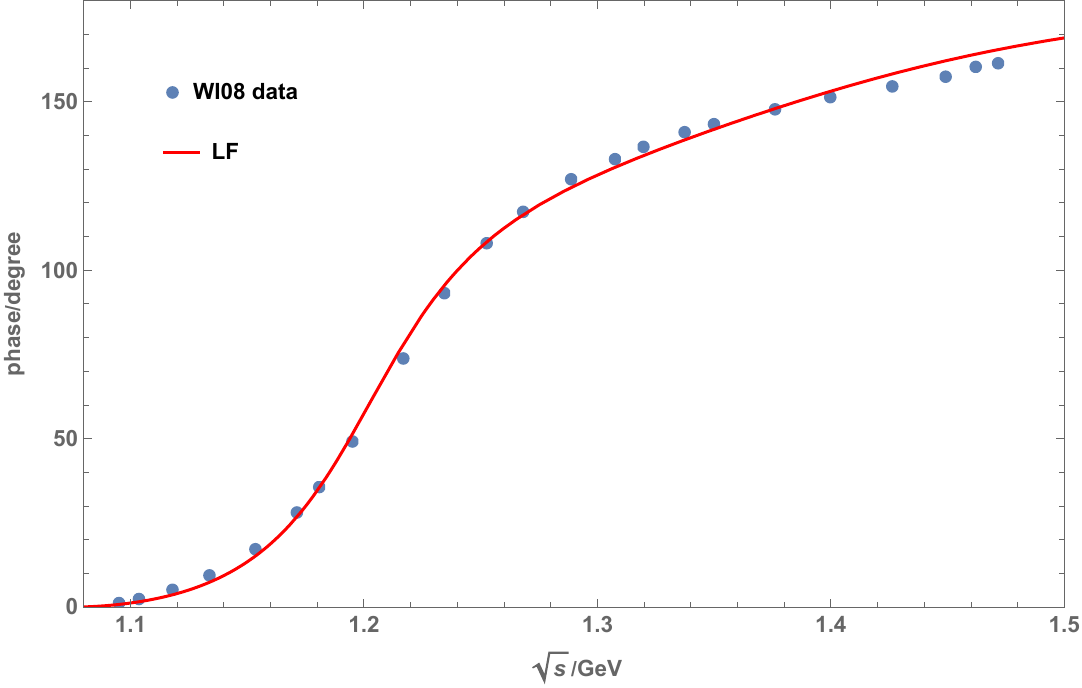}
\includegraphics[width=5cm]{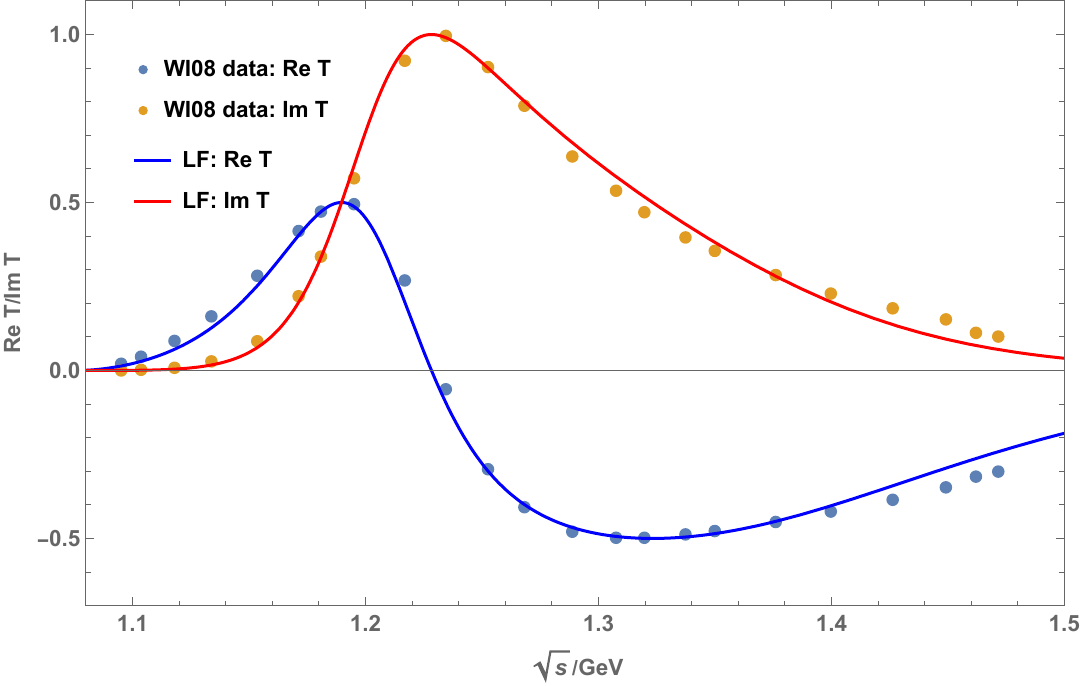}
\includegraphics[width=3.4cm]{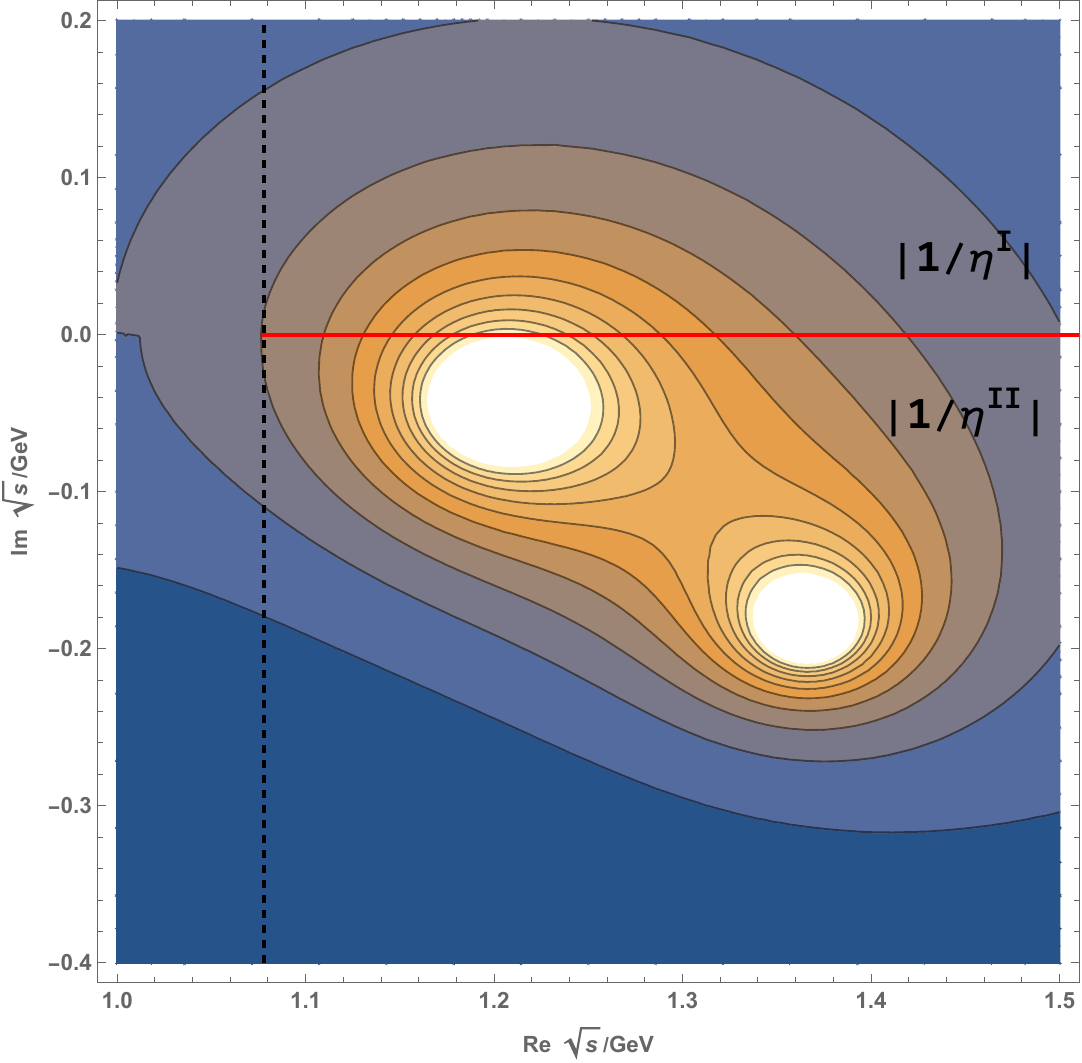}
	\caption{The real part and imaginary part of $T$ in $\pi p$ scattering~\cite{Arndt:2006bf}, and  fit results in the LF parametrization.}
	\label{fig-fit-delta-WI08}
\end{figure}

\subsection{$K^*_0(700)$ and $K^*_0(1430)$  \label{kSection}}

In the traditional framework, to distinguish whether a resonance peak, such as the $\rho$ or $\Delta(1232)$, is produced by a single pole in conjunction with a background or arises from two distinct poles has been a challenging task. Our subsequent study suggests that the $\pi K$ scattering process serves as an illustrative example showcasing a two-pole structure and the efficacy of the LF parameterization. The status of the $K_0^*(700)$ resonance has been a subject of intense debate over the past two decades, primarily due to its atypical resonance behavior and the difficulty in isolating it from background contributions when employing conventional Breit-Wigner parameterizations. Currently, the employment of model-independent approaches as in refs.~\cite{Descotes-Genon:2006sdr,Danilkin:2020pak,Pelaez:2020uiw}  has confirmed the existence of the $K_0^*(700)$ resonance, leading to a consensus within the scientific community.

In this study, we fit the magnitude $|a_0|$ and phase data of the $I=1/2$ $S$-wave $K\pi$ scattering amplitude obtained from the LASS data~\cite{Aston:1987ir} with the LF parametrization. We found that only a $(1,1)$ case without background contribution could obtain a reasonable fit of $|a_0|$ and $\delta_{\frac{1}{2}0}$ data below 1.6 GeV simultaneously. The higher $K\eta$ channel is not important {{when the $SU(3)$ symmetry is assumed and it is verified in many analyses, as depicted in ref.~\cite{Jamin:2000wn}.}} So it is expected that only the $\pi K$ channel is important. The fitted parameter values are listed in Table~\ref{tab-res-kstar}, where the extracted pole positions of  $K^*_0$'s resonances are also listed. For a direct perception about the pole positions,  the contour plot of $|1/\eta|$ in complex $\sqrt{s}$ plane at the fitted parameters are shown in Figure~\ref{fig-add-kstar-fit}.

\begin{figure}[htbp]
\centering
\includegraphics[width=5cm]{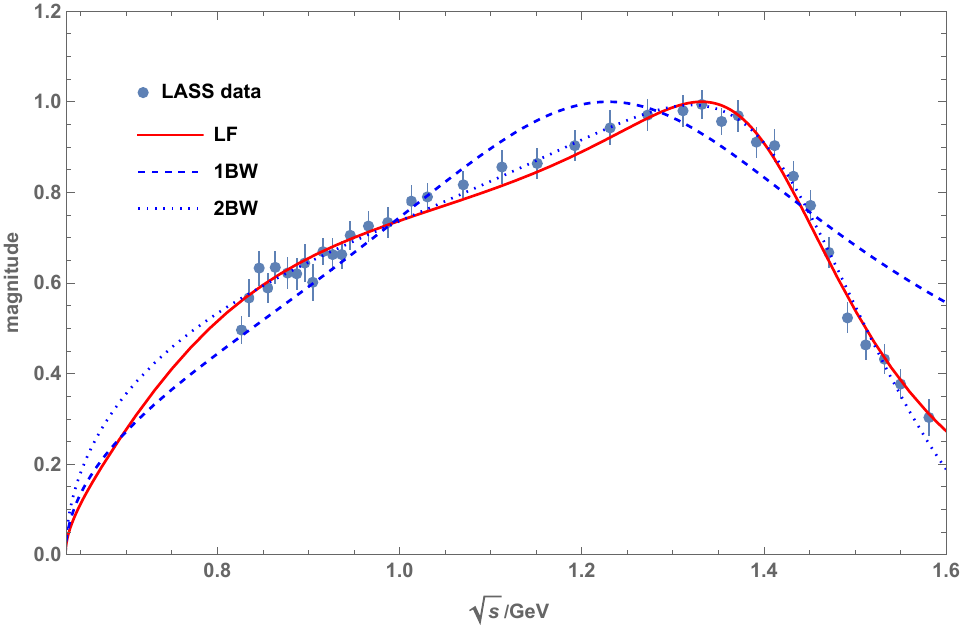}
\includegraphics[width=5cm]{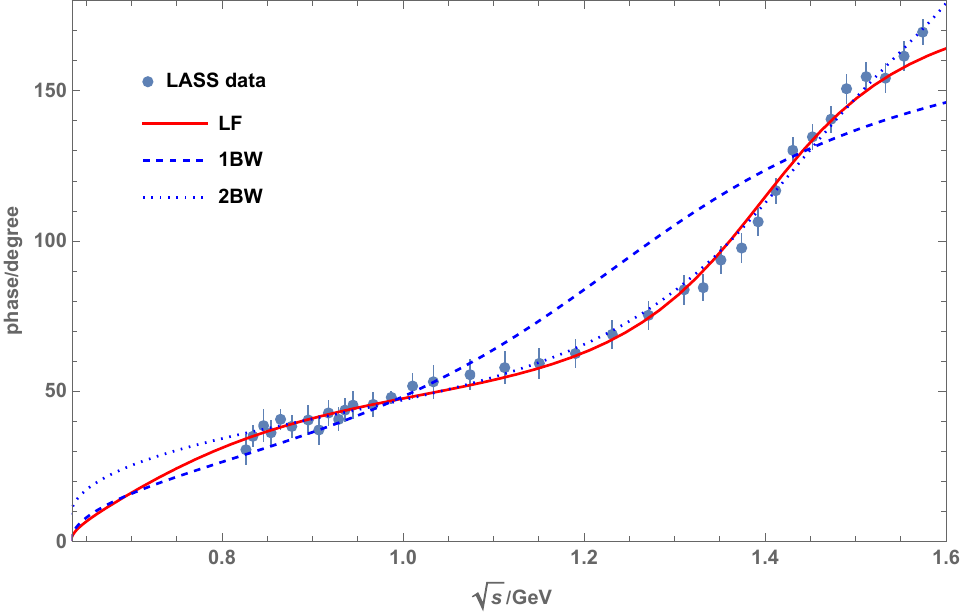}
\includegraphics[width=4cm]{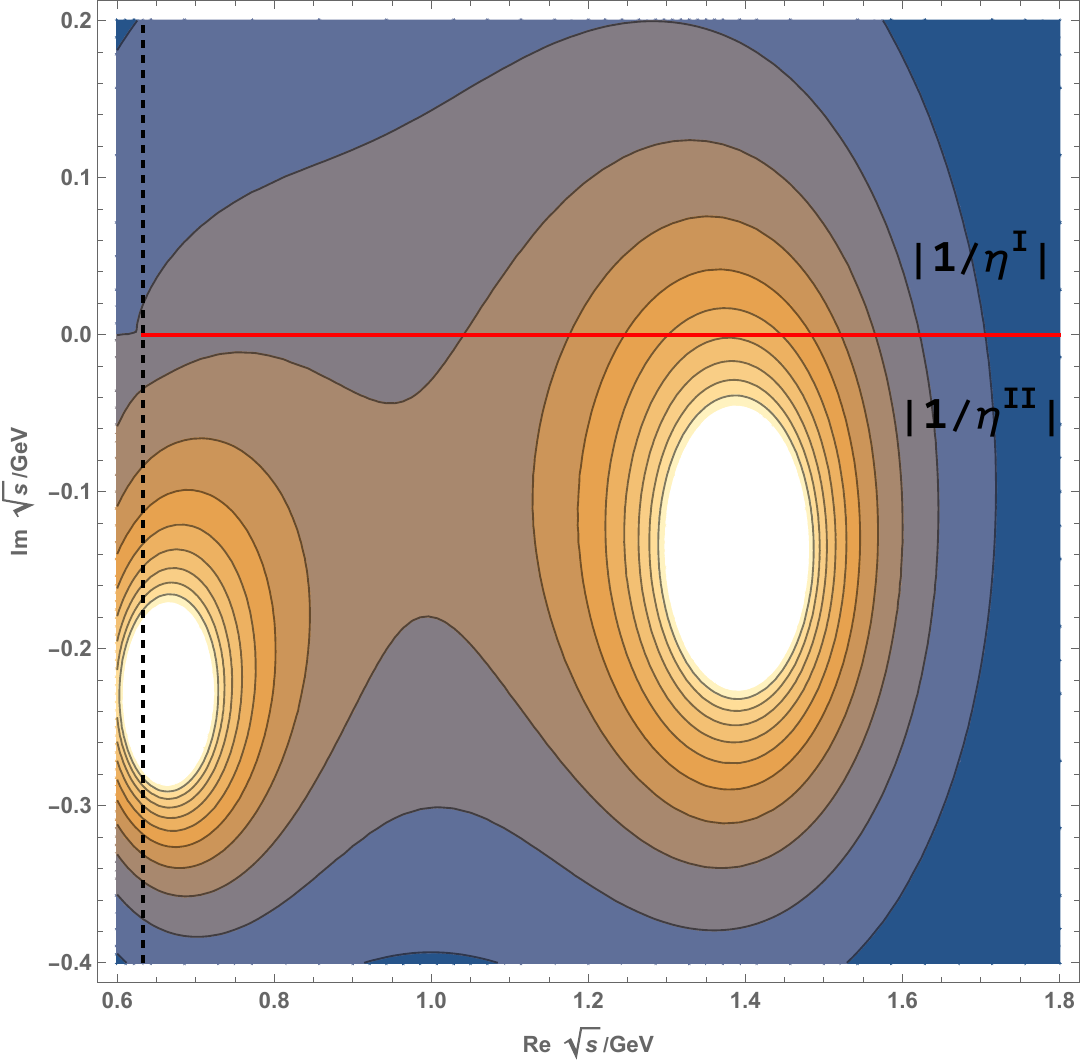}
\caption{The fit results for phase shift (left) and $|a_0|$ (middle), and the contour plot in the LF parametrization(right).}
	\label{fig-add-kstar-fit}
\end{figure}

\begin{table}[htbp!]
	\caption{Fit results for $K_0^*$ resonances.}
	\vspace{0.2cm}
	\label{tab-res-kstar}
	\centering
	\footnotesize
	\begin{tabular}{c|cccccc|ccc}
		\hline\hline
		Fit  & $\chi^2/dof$ & $m$ & $k_0$ & $g$ & $a_i$ & $\phi_i$ & $\sqrt{s_{R1}^{\Rmnum{2}}}$ (MeV) & $\sqrt{s_{R2}^{\Rmnum{2}}}$ (MeV) &$M_{BW}-i\Gamma_{BW}/2$(MeV)  \\
		\hline
LF  & 0.67 &1.28(2) & 0.43(3) & 1.92(19) & -&- & $658^{+27}_{-33}-i235^{+75}_{-85}$ & $1396^{+81}_{-60}-i141^{+71}_{-61}$ &- \\
\hline
1BW & 8.95 & 1.23(3) & - & 1.13(7) & - & - & - & - &  $(1229\pm28)-i(263\pm35)$ \\
\hline
		2BW$_1$ & 0.66 & 34(382) & - & 8(50) & 0.9(3.2) & -0.2(1.3) &- & - & $[(34\pm 382)-i (15\pm376)]\times10^3$ \\
		2BW$_2$ & - & 1.42(5) & - & 0.89(17) & 1.06(22) & 1.07(32) &- & - & $(1423\pm47)-i(171\pm66)$ \\
		\hline\hline
	\end{tabular}
\end{table}

In this study, it is noteworthy that the $\chi^2/\text{dof}$ value of the fit stands at approximately 0.67, indicating the high quality of the LF parameterization. Moreover, the two poles identified through the dressed propagator are consistent with the pole positions associated with the $K_0^*(700)$ and $K^*_0(1430)$ resonances.
The first broad resonance pole is situated at $(658\pm3)-\frac{i}{2}(470\pm20)$ MeV, closely resembling the PDG average values for $K^*_0(700)$, ranging in $(630-730)-i(260-340)$ MeV. At the same time, the second resonance pole, positioned around $1396^{+81}_{-60}-\frac{i}{2}282^{+142}_{-122}$ MeV, aligns closely with the PDG figures for $K^*_0(1430)$, specified as $(1431\pm6)-\frac{i}{2}(220\pm38)$ MeV. {{The bare mass of the $K^*_0$ is recorded at about 1.281 GeV, although it does not directly correspond to the poles' positions of $K_0^*(700)$ and $K^*_0(1430)$.  In the LF parametrization, this indicates that the bare $u\bar s$ $0^{++}$ state interacts with the $\pi K$ channel, resulting in two poles situated away from the bare mass.
It is interesting to observe that the analysis of  chiral perturbation theory with resonance, which is totally different from the present work, also prefers a bare mass at about $m_{K_0^*}=1.29$ GeV when investigating $S$-wave $\pi K$ scattering~(see~\cite{Jamin:2000wn}).}}

In our comparative analysis, we conducted fits using the Breit-Wigner (BW) parameterization. Notably, the fit employing a single BW form failed to capture both the phase shift and magnitude data adequately, as evidenced by the discrepancy shown by the dotted line in Figure~\ref{fig-add-kstar-fit}.
In a separate fit employing two BW forms, a total of eight parameters were necessitated. These additional parameters, $a_i$s and $\phi_i$s, were crucial for delineating the magnitude and relative phase of each BW resonance. However, the outcome of this fit unveiled a remarkably broad resonance characterized by an excessively large width, rendering it challenging to discern from the background signal.

{{We also made a trial fit with the phase space factor $\frac{E_1E_2k}{E}$ replaced by $\frac{k}{E}$, which is similar to the analysis of Tornqvist~\cite{Tornqvist:1995kr}, no $K_0^*(700)$ but a subthreshold bound state was found. This might be the reason why the $K_0^*(700)$  was not reported in ref.~\cite{Tornqvist:1995kr}. }}

The LF parameterization provides a natural rationale for why $K^*_0(700)$ and $K^*_0(1430)$ collectively contribute to a total phase shift of approximately $180^\circ$, as evidenced in the $IJ=\frac{1}{2}0$ $\pi K$ scattering phase around the energy region of 1.6 GeV. {{Notably, the Sill distribution, which parameterizes $K^*_0(700)$ and $K^*_0(1430)$ in two distinct forms, fails to describe the $\pi K$ spectral function well~\cite{Giacosa:2021mbz}, which also suggest that the two states are more appropriately parameterized together, as proposed in our approach.}

\subsection{$f_0$ states \label{f0Section}}
In the scenarios discussed earlier where  a single discrete bare state couples predominantly to a single decaying channel, the LF parameterization and BW parameterization for multichannel systems can be utilized for fitting data, especially when the coupling is weak.
However, in the cases characterized by strong couplings, the approach may differ. In standard data fitting procedures employing the BW parameterization, each lineshape peak typically requires to the introduction of a BW resonance. In contrast, in the fits utilizing the LF parameterization, dynamically generated poles can play a significant role and potentially manifest as isolated lineshape peaks. This feature complicates the interpretation of the  results, as the presence of such dynamically generated poles may challenge traditional analytical interpretations and assertions.

The classification of light $f_0$ states presents an intriguing case study to showcase distinctive features. Within the PDG table, several $f_0$ states are identified below 2.0 GeV, including $f_0(500)$, $f_0(980)$, $f_0(1370)$, $f_0(1500)$, and $f_0(1710)$, derived through various analyses. These states are expected to contribute to the isoscalar $S$-wave $\pi\pi$ scatterings.

Here, we tentatively apply the $(2,2)$ case of the LF parameterization to fit two datasets: the phase shift of $IJ=00$ $\pi\pi$ scattering and intensity data from $J/\psi\to\gamma\pi^0\pi^0$ decay. In this analysis, we would demonstrate that $f_0(500)$ and $f_0(1370)$ may form a two-pole structure emerging from a single bare state involving a $u\bar u$ seed. Simultaneously, we suggest that $f_0(980)$ and $f_0(1710)$ could originate from the $s\bar{s}$ bare state, with strong couplings to the $\pi\pi$ and $K\bar K$ channels.
It's important to note that the intention here is not to comprehensively unravel the intricate nature of these $f_0$ states and their implications in various decay processes and scattering phenomena, as these topics have been extensively discussed in the literature~\cite{Zhou:2020moj,Zhou:2020vnz,Eichten:1979ms, vanBeveren:1983td, Tornqvist:1995kr, Li:2009pw, Limphirat:2013jga, Achasov:2012ss, Zhang:2009gy, Segovia:2011zza, Wolkanowski:2015lsa, Wolkanowski:2015jtc, Yao:2020bxx}. Here, our primary aim is to assess whether the LF parametrization method can effectively reproduce the provided data or not.

We first fit the phase shift data of $IJ=00$ $\pi\pi$ scattering, referring to works like \cite{Estabrooks:1974vu,Protopopescu:1973sh}, with eight parameters. The fit results, including the values of these parameters, are displayed in Table~\ref{tab-res-f0}. The eight parameters determined in the fitting process consist of two bare masses, $m_a$ and $m_b$, four coupling strengths denoted as $g_{\alpha i}$, and two suppression factors $k_{0,i}$.
Table~\ref{tab-res-f0} presents four poles situated close to the physical region, with the associated contour plot of $\eta^{i}$ on the $i$-th sheets illustrated in Figure~\ref{fig-f0-fit}. These poles are proximate to $f_0(500)$, $f_0(980)$, $f_0(1370)$, and $f_0(1710)$. Notably, the mass of the highest pole is larger than  that of $f_0(1710)$. One possible explanation for this observation is that the smooth phase shift above 1500 MeV may poses challenges in constraining the contribution of the highest pole effectively.

In the subsequent step, we proceed to fit the intensity data derived from the $J/\psi\to\gamma\pi^0\pi^0$ decay \cite{BESIII:2015rug} within the energy range of (0.275, 1.9)~GeV/c. To account for potential background contributions, we introduce two additional parameters into the fit. The spectral function is expressed as
\bqa
events\propto |T_{J/\psi\to\gamma\pi^0\pi^0}|^2=|\sum_{\alpha,\beta=a}^b\pi g_\alpha[\eta^{-1}]_{\alpha\beta}f_{\beta 1}|^2,
\eqa
 Here, $g_a$ and $g_b$ serve as free parameters determined within the fitting process, representing the coupling of the resonances $a$ and $b$ with the decay process $J/\psi\to\gamma a/b$.
We assume that the coupling through electroweak interactions varies mildly with system energy compared to strong interactions. Consequently, we parameterize the $J/\psi\to\gamma a/b$ couplings as two parameters, $g_a$ and $g_b$, to accommodate this simplified model.

\begin{table}[htbp!]
	\caption{Fit results for $f_0$'s resonances from $\pi\pi$ phase shift and amplitude intensity data.
		}
	\vspace{0.2cm}
	\label{tab-res-f0}
	\centering
	\tiny
	\begin{adjustbox}{angle=0}
	\begin{tabular}{ccc|ccccccccccccc|cccc}
\hline\hline
\multicolumn{3}{c}{Fit}  & $\chi^2/dof$ & $m_a$ & $m_b$ & $k_{0,1}$ & $k_{0,1}$ & $g_{a1}$ & $g_{a2}$ & $g_{b1}$ & $g_{b2}$ & $g_a$ & $g_b$ &  &  & $\sqrt{s_{R1}^{\Rmnum{2}}}$/MeV  & $\sqrt{s_{R2}^{\Rmnum{2}}}$/MeV & $\sqrt{s_{R1}^{\Rmnum{3}}}$/MeV  & $\sqrt{s_{R2}^{\Rmnum{3}}}$/MeV   \\
\hline	
			& 	& phase  & 0.48 & 1.29(3) & 1.76(9) & 0.65(2) & 0.37(8) & 1.21(20)& $-$0.73(22)& $-$2.11(12)& $-$4.16(1.22)& - & - & & & 479$-i$265 & 983$-i$19  & 1346$-i$266 & 2059$-i$156 \\
\cline{2-20}
		 	& 	& sol1  & 2.27 & 1.166(6)& 1.552(2)& 0.500(7)& 0.331(7)& 1.50(2) & $-$3.57(11)& $-$2.00(3) & 1.56(4)      & 10.0(4) & 55.4(7) &  &  & 380$-i$166 & 949$-i$206 & 1430$-i$58 & 1740$-i$105 \\

		 	& 	& sol2  & 1.69 & 1.107(9)& 1.622(7)& 0.337(9)& 0.383(5)& 2.04(6) & $-$2.98(6) & $-$1.49(5) & 2.09(4)      & 39(5)   & 193(20) &  &  & 320$-i$203 & 844$-i$140 & 1418$-i$85 & 1749$-i$60 \\

\hline\hline
	\end{tabular}
\end{adjustbox}
\end{table}

\begin{figure}[htbp]
\centering

\includegraphics[width=5cm]{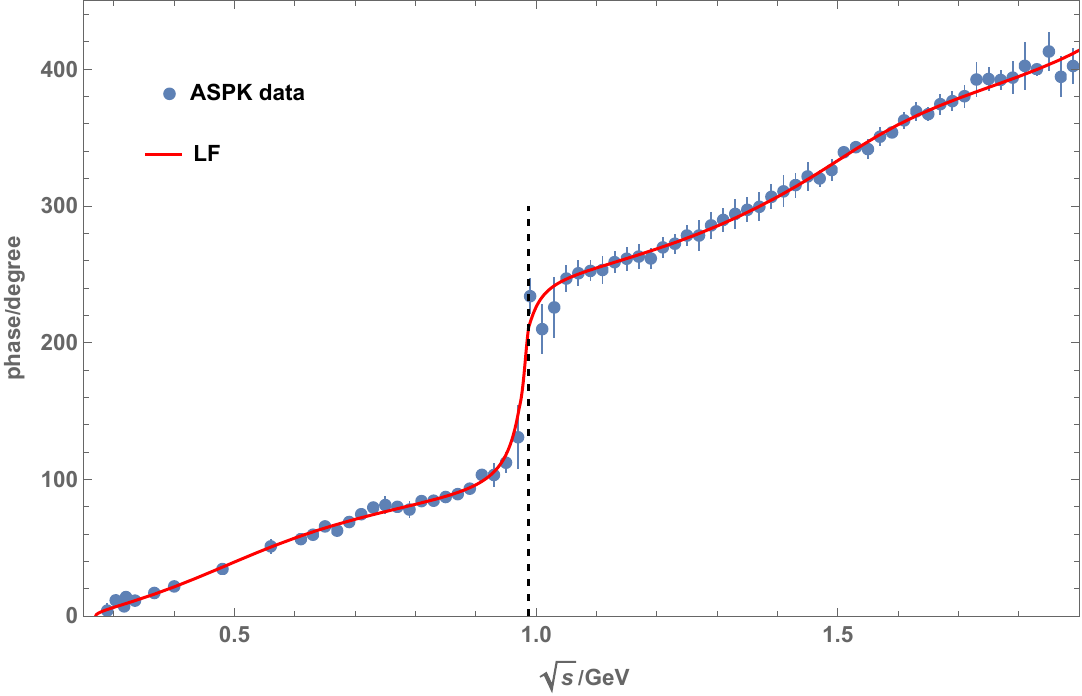}
\includegraphics[width=5cm]{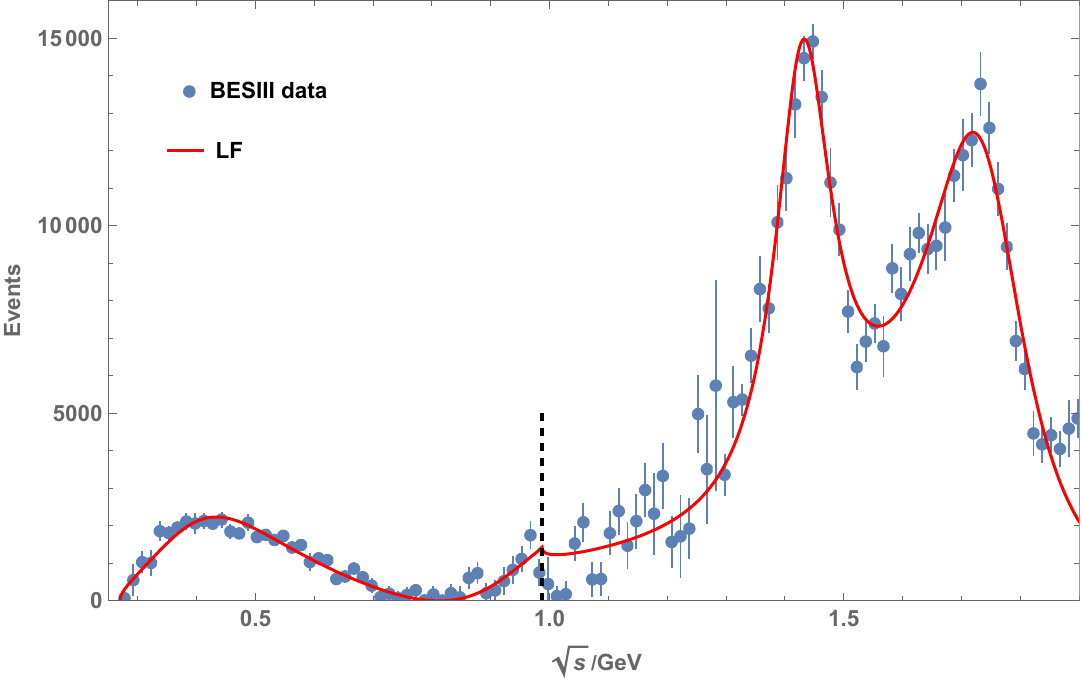}
\includegraphics[width=5cm]{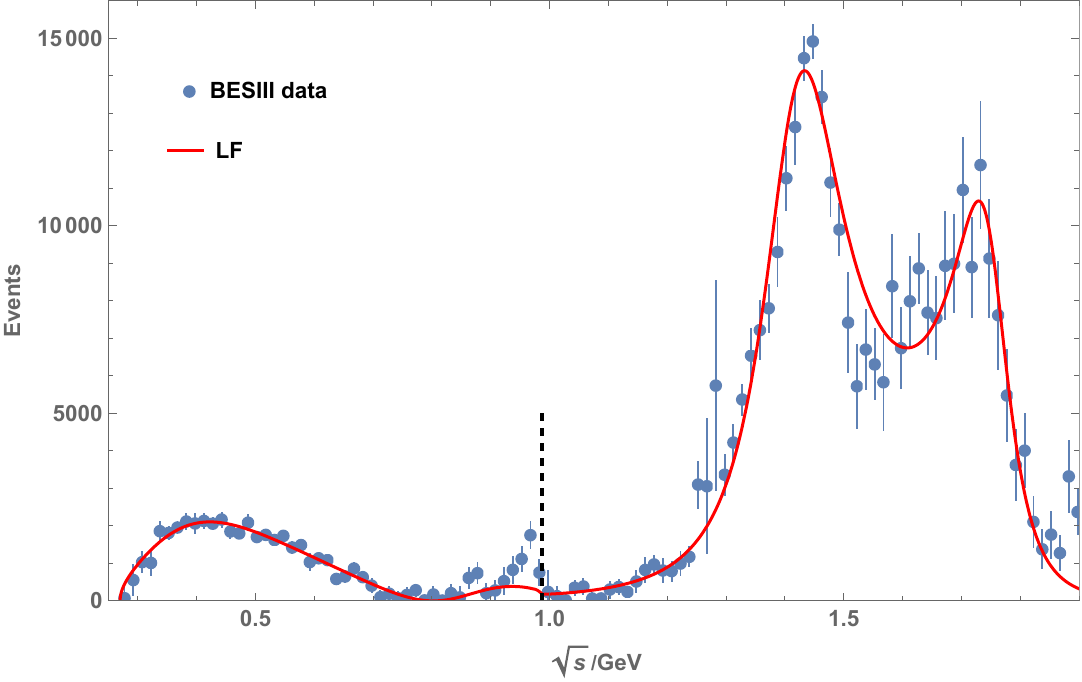}

\includegraphics[width=5cm]{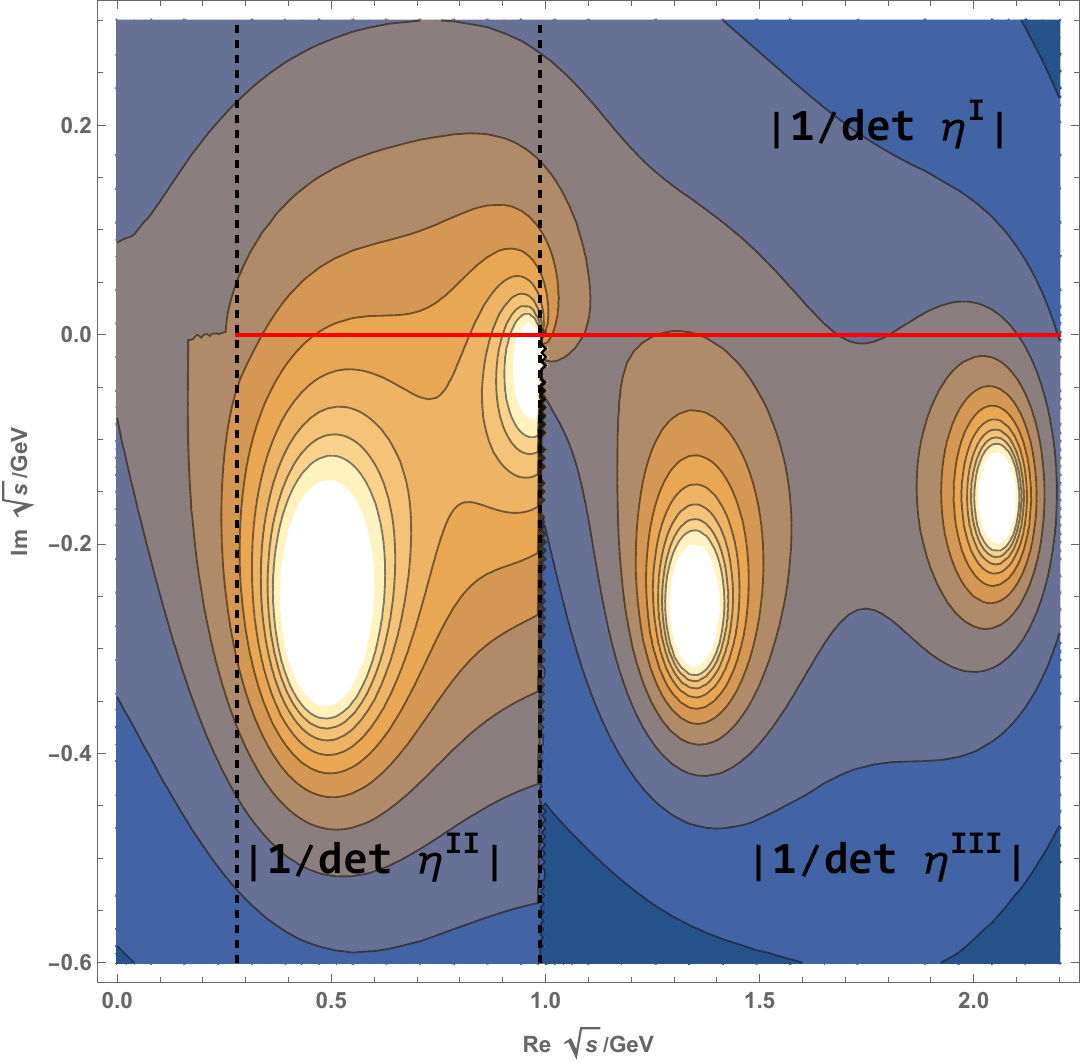}
\includegraphics[width=5cm]{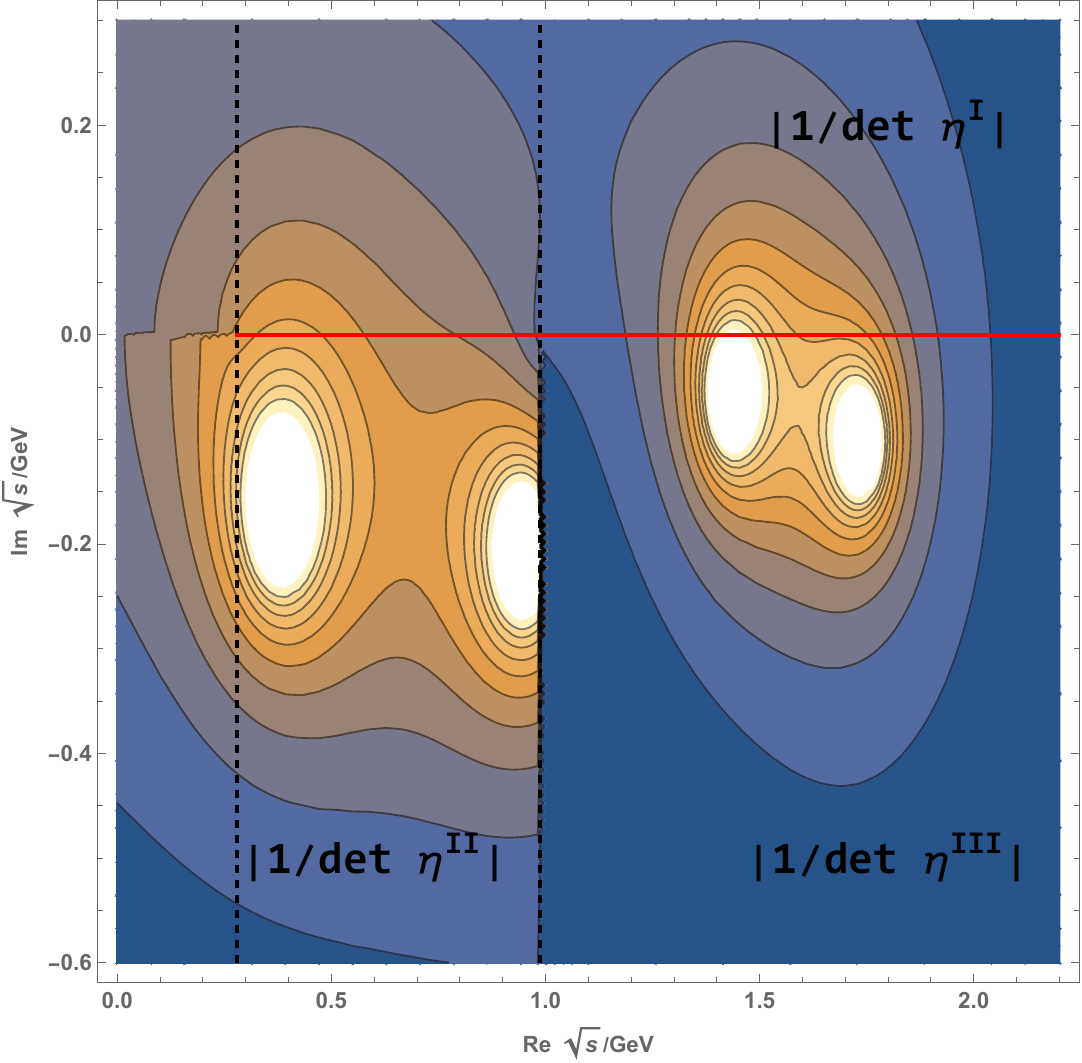}
\includegraphics[width=5cm]{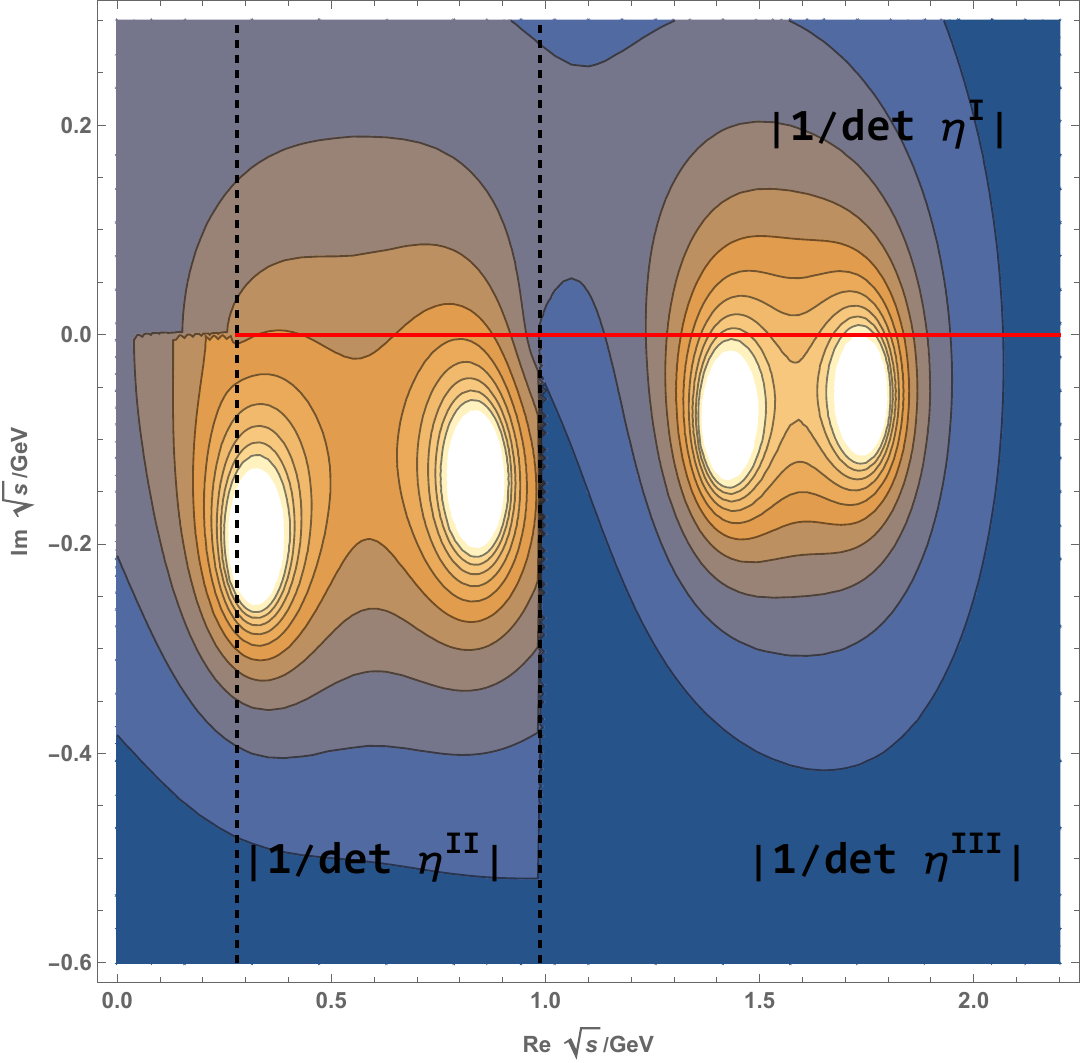}

\caption{The fit and contour plots for the LF parameterization of $\pi\pi$ scattering phase data~(left) and the solution 1~(middle) and 2~(right) of event data  of BESIII in $J/\psi\to\gamma\pi\pi$~\cite{BESIII:2015rug}, respectively.}
	\label{fig-f0-fit}
\end{figure}

In the analysis of event data solutions from the $J/\psi\to\gamma\pi\pi$ process by BESIII, the overall distribution of four lineshape peaks is aptly captured, although with limitations in reproducing the characteristics of the $f_0(980)$ peak. {{ In a simplified study postulating that a $s\bar{s}$ bare state couples to the $K\bar{K}$ channel~\cite{Zhou:2020moj}, the $f_0(980)$ could emerges as a dynamically-generated bound state residing below the $K\bar{K}$ threshold. When the lower $\pi\pi$ threshold is also considered as studied here, the bound state will transit to a second-sheet pole. However, it seems to be difficult to balance the determination of the pole position and description of the events data in a large energy region up to 1.9 GeV.}}
Another possibility of the discrepancy might be  that the presence of background contributions plays a significant role in shaping the event data extraction. Additionally, the impact of coupling channel effects, particularly involving the $\eta^{(')}\eta^{(')}$ thresholds, may also be crucial in this context. Given the complexities inherent in this dataset, our current study does not aim to provide an exhaustive fit, recognizing the multifaceted factors at play in this analysis.

\section{Discussions and Summary  \label{summarySection} }

This study is devoted to the exploration of utilizing the extended Lee-Friedrichs model as a basis of the parameterization method of the amplitude. The rigorous solutions of LF model reveal that this parameterization of scattering amplitude exhibits the  analyticity structure required by scattering theory. We verified its applicability on several cases involving different numbers of discrete states and decay channels, including the details on scattering amplitudes, distribution functions, dressed propagators, and the positions of their poles.

In the simplest scenario involving only one discrete state and one decaying channel, which is denoted as the $(1,1)$ case, the LF parametrization method introduces an energy-dependent tail to the distribution function. When the coupling strength is small, the distribution function behaves similarly to the improved BW parametrization approach, offering a comparable fit quality for resonance structures featuring a single lineshape peak. However, the interpretation of the LF parametrization diverges, primarily due to its analyticity structure enabling the straightforward extraction of poles on unphysical Riemann sheets.

Generally, there could be dynamically generated poles in LF parametrization which may lie far way when the coupling is small. In the cases where the coupling increases to the extent such that the dynamically generated poles come close to the physical region, these two poles contribute distinct observable lineshape peaks, and may often be identified as two distinct resonances. The existence of these dynamically generated poles may be  typically attributed to an exotic origin or molecular origin.

In the instance involving one discrete state and two decaying channels, the parametrization formula and its behavior may be simplified to the commonly used Flatt\'{e} parametrization form. In the cases when the resonance pole is close to the threshold, the threshold triggers a cusp effect, incising the lineshape peak in the lower channel and resulting in an asymmetric lineshape peak.

For more general scenarios encompassing multiple discrete states and multiple scattering channels, the LF parametrization proves to be effective in fitting processes with strong coupled-channel effects. This  parametrization demonstrates its advantages for comparative analyses against the $K$-matrix parametrization.

In each scenario both relativistic and non-relativistic forms are available. While the relativistic form is preferred  for experimental analyses, the evaluation of the dispersion integral requires numerical computation. In contrast, the non-relativistic scenario has a computational advantage, as the dispersive integration  can be performed and be expressed in terms of  the incomplete Gamma function, leading to substantial savings in computational time during fitting processes and pole extraction.

In this illustrative study, we present selected examples rather than an exhaustive analysis of each state. Notably, single-peak resonances like the $\rho(770)$ and $\Delta(1232)$ are effectively characterized by the LF parametrization, which captures the behavior of the high-energy tails. The analytical structure reveals that these peaks might be dominantly contributed by two poles. Particularly for asymmetric lineshapes such as the $\Delta(1232)$, the broader pole may reside in proximity to the physical region, whose existence need further theoretical explorations.
The $K_0^*(700)$ and $K_0^*(1430)$ exemplify a typical two-pole structure, where both poles exhibit customary lineshapes and comply with  the phase shift sum rule. Furthermore, the analysis of $f_0$ states serves as an illustration of the cases involving multiple discrete states and scattering channels, although a more intricate analysis is warranted.

To deepen our understanding, it is recommended that additional trial analyses be conducted in experimental investigations, comparing the outcomes of LF parametrization against BW parametrization. Such comparative analyses can shed light on the efficacy and subtleties of these parametrization methods in diverse experimental settings. {{For example,  in a recent analysis of $B^+\rightarrow D^{*\pm}D^{\mp}K^+$ by LHCb~\cite{LHCb:2024vfz}, a new charmonium-like state $\chi_{c1}(4012)$ was reported alongside the well-known $\chi_{c1}(3872)$ when analyzing the data using several parameterization tests. Prior to this observation, there had been theoretical conjectures that $\chi_{c1}(3872)$ is dynamically generated by the coupling of the bare $\chi_{c1}(2P)$ state, a long sought-after charmonium state,  to the $D\bar{D}^*$ channel~\cite{Zhou:2017dwj,Deng:2023mza}. Interpreting the $\chi_{c1}(4012)$ and $\chi_{c1}(3872)$ as a two-pole structure described within the LF parameterization, with the line shapes of $J\psi\pi\pi$ and $D\bar{D}^*$ channels illustrated in Figure \ref{fig-shape-X3872} (with several parameters set by hand for illustration), might signicantly reduce the fit parameters. Such an approach could be valuable in practice for the real data analysis and might be helpful in revealing the nature of these two states. }}

\begin{figure}[htbp]
	\centering
	\includegraphics[width=7cm]{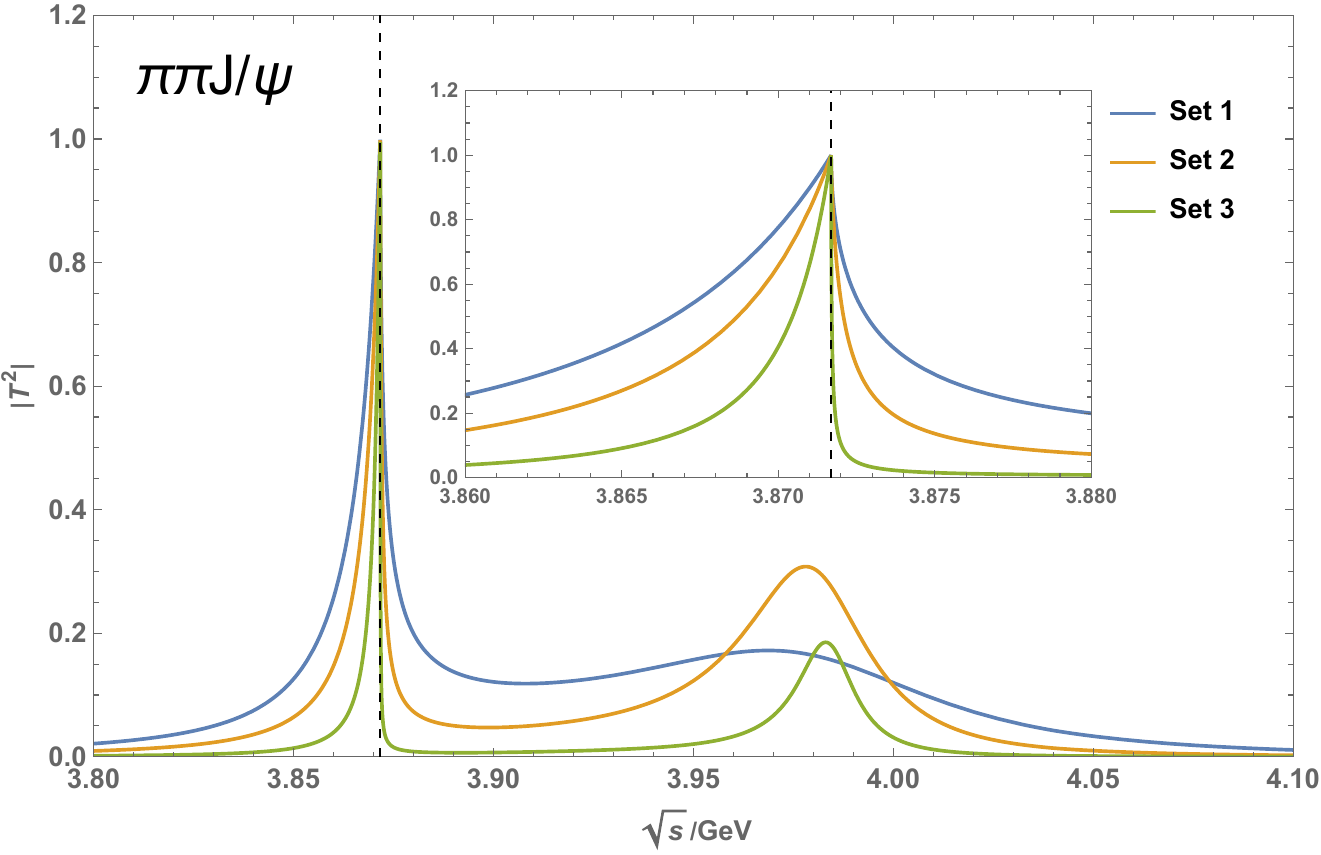}
	\includegraphics[width=7cm]{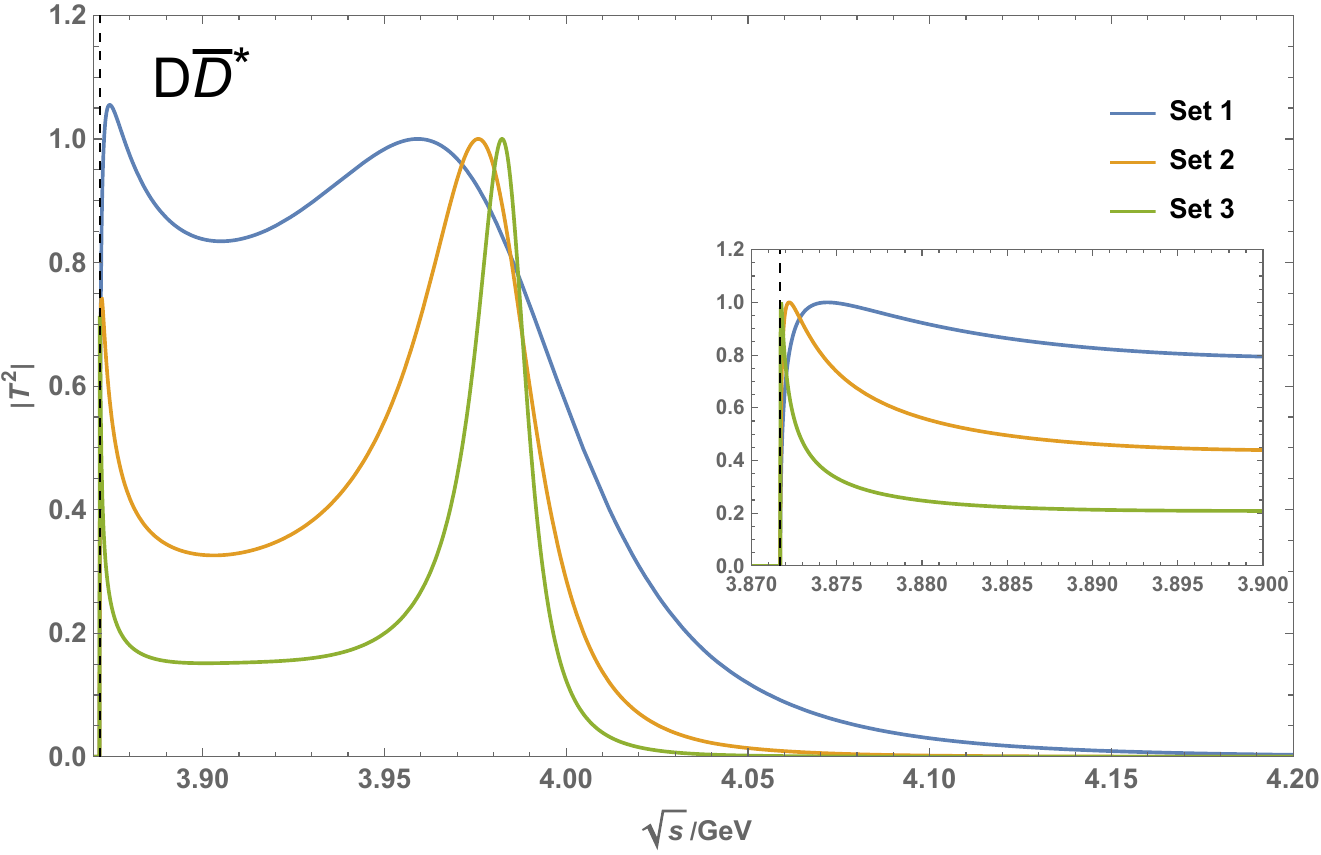}
	\caption{The $|T_{\pi\pi J/\psi}^2|$ and $|T_{DD^*}^2|$ lineshapes with three sets of parameters: (1) $m=3.95$, $g_1=0.3$, $g_2=1.0$, $k_{0,1}=k_{0,2}=0.3$; (2) $m=3.945$, $g_1=0.5$, $g_2=0.8$, $k_{0,1}=k_{0,2}=0.4$; and (3) $m=3.955$, $g_1=0.4$, $g_2=1.2$, $k_{0,1}=k_{0,2}=0.25$. The right plots are lineshapes in zoomed in region. The functions are normalized to the maximum. The dashed line shows the $DD^*$ threshold.}
	\label{fig-shape-X3872}
\end{figure}

Comparisons drawn between the BW parametrization and the LF parametrization reveal several key distinctions:

(1) In scenarios where the two-pole structure manifests as a single peak, the improved BW distribution may approximate the LF parametrization, although the BW parametrization fails to reflect the underlying pole structure. Strong coupling effects, where dynamic poles significantly affect observables, are better captured by the LF form, showcasing a distinct two-pole structure.

(2) When the pole is situated in proximity to the threshold, the LF parametrization is capable of transitioning into the Flatt\'{e} parametrization. Furthermore, when coupled channel effects are significant, parallels can be drawn between the LF parametrization and the $K$-matrix approach.

(3)The mass parameter $m$ in the LF parametrization may possess a more physical interpretation compared to the BW mass parameter. $m$ can be viewed as the quenched bare mass of the discrete state which is obtained from the quenched quark model. The bare discrete state couples to the scattering channels and generates the lineshapes of the observed data.

In summary, this paper introduces a  parameterization method grounded on the extended LF model. Through several fitting examples, we demonstrate the effectiveness of this method in experimental analyses and resonance parameter extraction. A crucial insight emerges, emphasizing how a two-pole structure can arise from the coupling of  a single bare state to the scattering channels through this parameterization. This potentially offers profound insights into the nature of resonance spectra, especially pertaining to exotic states.

\begin{acknowledgments}
 This work is supported by China National Natural Science Foundation
under contract  No. 12375132, No. 11975075, No. 12375078, No. 11575177, No.11947301, and No.12335002.
This work is also supported by ``the Fundamental Research Funds for the Central Universities".
\end{acknowledgments}

\bibliographystyle{apsrev4-1}
\bibliography{Ref}


\clearpage

\end{document}